%% file: accepted.tex
\documentclass[iop,apjl,tighten]{emulateapj}
\usepackage{amssymb}
\usepackage{amsmath}
\usepackage{times}
\usepackage{epsfig}
\usepackage{graphicx}
\usepackage{natbib}
\usepackage{color}
\usepackage{enumerate}
\include{def_bibtex}

\usepackage{arydshln}
\input{epsf}
\usepackage[T1]{fontenc}

\begin{document}
% \date{}

\title{Polarization modulation from Lense-Thirring precession in X-ray binaries}
\shorttitle{Polarization modulation}
\shortauthors{Ingram et al.}
\author{Adam Ingram\altaffilmark{1}, Thomas J. Maccarone
  \altaffilmark{2}, Juri Poutanen \altaffilmark{3}, \& Henric
  Krawczynski \altaffilmark{4}} 
\affil{$^1$Anton Pannekoek Institute, University of Amsterdam, Science
  Park 904, 1098 XH Amsterdam, the Netherlands, a.r.ingram@uva.nl; \\
$^2$Department of Physics, Texas Tech University, Box 41051, Lubbock,
TX 79409-1051, USA; \\
$^3$ Tuorla Observatory, Department of Physics and Astronomy, 
University of Turku, V\"ais\"al\"antie 20, FI-21500 Piikki\"o,
Finland; \\
$^4$ Physics Department and McDonnell Center for the Space Sciences, Washington University in St. Louis, 1 Brookings Drive,
CB 1105, St. Louis, MO 63130, USA.}

\submitted{Accepted for pubilcation in ApJ}

\begin{abstract} 
\noindent It has long been recognised that quasi-periodic oscillations
(QPOs) in the X-ray light curves of accreting black hole and neutron
star binaries have the potential to be powerful diagnostics of strong
field gravity. However, this potential cannot be fulfilled without a
working theoretical model, which has remained elusive. Perhaps the
most promising model associates the QPO with Lense-Thirring precession
of the inner accretion flow, with the changes in viewing angle and
Doppler boosting modulating the flux over the course of a precession
cycle. Here, we consider the polarization signature of a precessing
inner accretion flow. We use simple assumptions about the
Comptonization process generating the emitted spectrum and take all
relativistic effects into account, parallel transporting polarization
vectors towards the observer along null geodesics in the Kerr
metric. We find that both the degree of linear polarization and the
polarization angle should be modulated on the QPO frequency. We
calculate the predicted absolute rms variability amplitude of the
polarization degree and angle for a specific model geometry. We find
that it should be possible to detect these modulations for a
reasonable fraction of parameter space with a future X-ray polarimeter
such as \textit{NASA's} \textit{Polarization Spectroscopic Telescope
  Array} (\textit{PolSTAR}: the satellite incarnation of the recent
balloon experiment \textit{X-Calibur}).
\end{abstract}


\section{Introduction}
\label{sec:intro}

Low frequency quasi-periodic oscillations (hereafter QPOs) with frequencies ranging from $\sim
0.1-30$ Hz are routinely observed in the X-ray light curves of black
hole (BH) and neutron star binary systems. The
QPO properties correlate strongly with spectral state, which
evolves from the hard power law dominated \textit{hard state} to the
disk blackbody dominated \textit{soft state} on timescales of
$\sim$months (\citealt{Tananbaum1972}; see \citealt{VDK2006};
\citealt{Done2007}; \citealt{Belloni2010} for reviews). The QPO
frequency increases as the spectrum softens and the amplitude rises to
a peak in the hard intermediate state (HIMS: \citealt{Belloni2010})
before reducing as spectral evolution further continues
(e.g. \citealt{Muno1999}; \citealt{Sobczak2000}). Since the flux also
peaks in this state, the HIMS is ideal for studying QPOs.

The disk blackbody spectral component is from a geometrically thin
accretion disk (\citealt{Shakura1973}; \citealt{Novikov1973}) and the
power law component results from Compton up-scattering of cool seed
photons by some cloud of hot electrons close to the BH
(\citealt{Thorne1975}; \citealt{Sunyaev1979}). Some fraction of these
seed photons are provided by the disk, with the rest generated
internally in the electron cloud via cyclo-synchotron radiation
(\citealt{Ghisellini1988}; \citealt{Poutanen2009};
\citealt{Veledina2011a}). The exact geometry of the Comptonizing cloud
is still a matter of debate but a prominent interpretation is the
\textit{truncated disk model} (\citealt{Ichimaru1977};
\citealt{Esin1997}; \citealt{Poutanen1997}; \citealt{Done2007};
\citealt{Gilfanov2010}), in which the geometrically thin disk
evaporates inside of some radius larger than the innermost stable
circular orbit (ISCO) to form a geometrically thick, optically thin
($\tau\sim 1$) accretion flow (hereafter the inner hot flow). The
spectral transitions can be explained if the truncation radius moves
smoothly from $R_o\sim 60 R_g$, where $R_g = GM/c^2$, in the hard
state to $R_o = R_{\rm ISCO}$ in the soft state.

It has long been recognised that QPOs have the potential to be
powerful diagnostics of the regions close to BHs. However, this
potential cannot be realised without a working model, which has long
remained elusive. Suggested QPO mechanisms in the literature either
consider particle orbits in General Relativity (GR) (\citealt{Stella1998};
\citealt{Wagoner2001}; \citealt{Schnittman2006}) or instabilities
in the accretion flow (\citealt{Tagger1999};
\citealt{Cabanac2010}). Perhaps the most successful model for
explaining the array of observational properties attributes the QPO to
the effect of frame dragging. In GR, a spinning
massive object twists up the surrounding space-time, leading to
precession of particle orbits inclined with the BH equatorial
plane. This is generally called Lense-Thirring precession after the
authors who originally derived it (\citealt{Lense1918}), although
their derivation was only in a weak field limit, coming well before
the derivation of the Kerr metric
(\citealt{Kerr1963}). \cite{Stella1998} first suggested that the QPO
results from Lense-Thirring precession, considering only the
precession frequency of test masses at different
radii. \cite{Schnittman2006} considered instead a precessing ring in
the accretion disk. However, phase resolved spectroscopy reveals that
it is the Comptonized spectral component which oscillates and not the
disk (\citealt{Markwardt1999}; \citealt{Revnivtsev2001a};
\citealt{Sobolewska2006}; \citealt{Axelsson2013}). \cite{Ingram2009}
suggested that the entire inner flow precesses as a solid body,
motivated by the General Relativistic Magneto-Hydrodynamic (GRMHD)
simulations of \cite{Fragile2007}. This
precessing flow model naturally explains the observed QPO spectrum,
predicts the correct range of frequencies for BHs
(\citealt{Ingram2009}), has been incorporated into a full model 
for the power spectral properties of BHs (\citealt{Ingram2011};
\citealt{Ingram2012}; \citealt{Ingram2013};
\citealt{Rapisarda2014}) and has been extended to also account for the
optical QPOs observed from BHs (\citealt{Veledina2013};
\citealt{Veledina2015}). The model can also explain the range of QPO
frequencies observed in low accretion rate neutron stars
(\citealt{Ingram2010}), but not the $11$ Hz pulsar in the globular
cluster Terzan 5 (\citealt{Altamirano2012}).

In this Paper, we consider the X-ray polarization signature from a
precessing inner flow. Photons emerging from a Comptonizing slab are
expected to be polarized, with the polarization degree as a function
of viewing angle depending on optical depth of the slab
(\citealt{Chandrasekhar1960}; \citealt{Rybicki1979}; \citealt{Loskutov1982};
\citealt{Sunyaev1985}; \citealt{Poutanen1996}). Although the
polarization degree is predicted to be less than $\sim 20\%$, the
current generation of proposed X-ray polarimetry missions should
comfortably be able to detect a signal. Polarimetry promises to
provide a powerful extra lever arm to interpret the X-ray signal from
BHs. The predicted energy dependence of the polarization signature has
been extensively explored (e.g. \citealt{Stark1977};
\citealt{Dovciak2008}; \citealt{Li2009}; \citealt{Schnittman2009};
\citealt{Schnittman2010}; \citealt{Henric2012}). Variability of the
signal is less well studied, although Lense-Thirring precession did
provide an early motivation for X-ray polarimetery in the 1970s (Knox
Long, private communication). More recently, \cite{Dovciak2008a} and
\cite{Zamaninasab2011} considered the polarization signature of
orbiting hotspots and \cite{Marin2015} considered the polarization
signatures of obscuring clouds, all for the case of Active Galactic
Nuclei. In this Paper, we predict a modulation in the polarization
signature from bright stellar mass BHs on a timescale of $\sim 1$
s. We use simple parameterizations for the angular dependence of
emission and polarization designed to mimic the results of
\citet[hereafter ST85]{Sunyaev1985} for the $\tau=1$ case. We then
take GR fully into account by ray tracing photon paths from the
precessing flow geometry to the observer, assuming the Kerr metric.

In Sections \ref{sec:geometry} and \ref{sec:method}, we outline our
assumed geometry and formalism for calculating the observed
polarization degree and angle as a function of QPO phase. In
Section \ref{sec:results}, we consider a specific flow geometry appropriate
for the HIMS and a QPO frequency of $\sim 1-2$ Hz. We calculate the
amplitude of the modulation in both polarization degree and angle for
the full range of observer positions. In Section \ref{sec:discussion},
we discuss the prospects of detecting this signature and suggest
improvements that can be made to our modelling assumptions in future.
Throughout this Paper, we express distance in units of $R_g$ and time
in units of $c/R_g$ in order to exploit the scale invariant nature of
GR. We use Einstein's summing convention with Greek indices taken to
run from 0 to 3 and Roman indices taken to run from 1 to 3. We adopt
the 4-vector formalism with negative time entries and positive spatial
entries.

\section{Geometry}
\label{sec:geometry}

We assume the geometry proposed by \cite{Ingram2009}, following
\cite{Ingram2012a} and \citet[hereafter VPI13]{Veledina2013}. Specifically, the BH and
binary system spin axes are misaligned by some modest angle
$\beta$. The outer thin disk is assumed to be in the binary plane and
the flow spin axis is assumed to precess around the BH spin axis,
always maintaining the same misalignment angle $\beta$. In this
section, we first describe our coordinate system followed by a
description of the assumed geometry of the inner flow.

\subsection{Coordinate system}
\label{sec:coord}

\begin{figure}
 \includegraphics[height=9.5cm,width=9.5cm,trim=11.5cm 9.0cm 10.0cm
 3.0cm,clip=true]{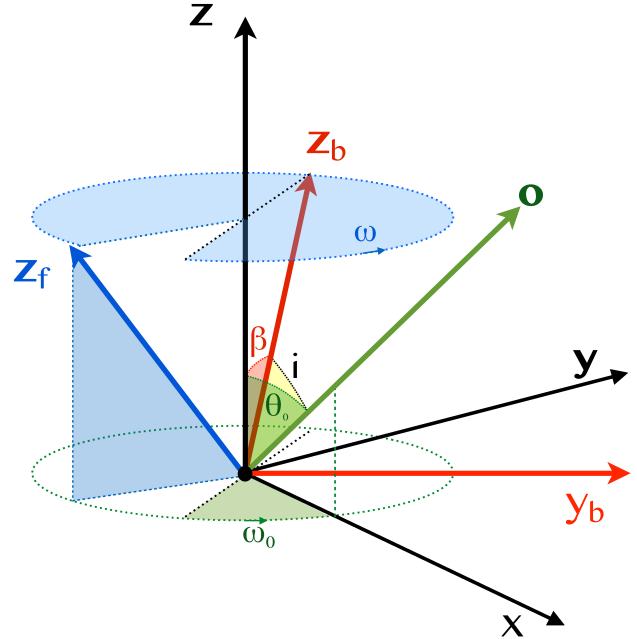}
\vspace{-3mm}
 \caption{Coordinate system used in this Paper. We define right handed
   Cartesian coordinate systems with z-axes aligned with the BH, flow
   and binary spin axes. The BH (black) and binary (red) z-axes are
   misaligned by the angle $\beta$ and the binary system x-axis is
   chosen to be in the unique plane shared by the BH and binary
   z-axes. The vector $\mathbf{\hat{o}}$ represents the observer's
   line of sight, and the BH x-axis is chosen as the projection of
   this vector onto the BH equatorial plane. The flow z-axis (blue)
   precesses around the BH z-axis with precession angle $\omega$,
   always maintaining the misalignment angle $\beta$. The flow and
   binary z-axes are therefore aligned when $\omega=180^\circ$ and
   misaligned by an angle $2\beta$ when $\omega=0^\circ$.}
 \label{fig:coord}
\end{figure}
%Left, Bottom, Right, Top

Figure \ref{fig:coord} illustrates the coordinate system used in this
paper. As in VPI13, we assume the binary spin axis is misaligned with
the BH spin axis by an angle $\beta$. We define a `binary' coordinate
system in which the z-axis, $\mathbf{\hat{z}_b}$, aligns with the
binary spin axis and therefore the plane of the binary is simply the
plane perpendicular to $\mathbf{\hat{z}_b}$. We choose to align the
binary system x-axis with the projection of the BH spin axis on the
plane of the binary. In binary coordinates, the vector pointing from
the BH to the distant observer is given by
\begin{equation}
\mathbf{\hat{o}} = ( \sin i \cos\Phi , \sin i \sin \Phi , \cos i ),
\end{equation}
where $i$ is the inclination angle and $\Phi$ is the viewer
azimuth. We use these angles to define the observer's position because
it is possible to measure $i$ via dynamical methods
(e.g. \citealt{Orosz2004, Orosz2011}). Throughout this Paper,
  a hat denotes a unit vector.

We assume that the flow spin axis ($\mathbf{\hat{z}_f}$) precesses
around the BH spin axis, maintaining a misalignment angle $\beta$ as
the precession angle $\omega$ increases. This means that the
misalignment between the flow and binary spin axes varies over a
precession cycle between $0$ and $2\beta$. We define a `flow'
coordinate system with exactly the same relationship to the binary
coordinate system as introduced in VPI13. Here, however, we will
perform calculations in the Kerr metric which is azimuthally symmetric
around the BH spin axis, in contrast to the Schwarzschild metric used
in VPI13 which is spherically symmetric. Consequently, our
calculations are greatly simplified by defining a `BH' coordinate
system with basis vectors $\mathbf{\hat{x}}$, $\mathbf{\hat{y}}$ and
$\mathbf{\hat{z}}$. The z-axis aligns with the BH spin axis, allowing
us to use Boyer-Lindquist coordinates, and the x-axis aligns with the
projection of $\mathbf{\hat{o}}$ onto the BH equatorial plane. In this
coordinate system, the observer's line of sight can be written as
\begin{equation}
\mathbf{\hat{o}} = ( \sin\theta_0 , 0 , \cos\theta_0 ),
\end{equation}
where
\begin{equation}
\cos\theta_0 = \sin i \cos\Phi \sin\beta + \cos i \cos\beta,
\end{equation}
is the cosine of angle between the observer's line of sight and the BH spin
axis. We can express the flow basis vectors in this coordinate system
as
\begin{eqnarray}
\mathbf{\hat{x}_f} & = & (-\cos\beta \cos(\omega-\omega_0) , -\cos\beta
\sin(\omega-\omega_0) , \sin\beta ) \nonumber \\
\mathbf{\hat{y}_f} & = & (\sin(\omega-\omega_0) , -\cos(\omega-\omega_0) , 0 ) \nonumber \\
\mathbf{\hat{z}_f} & = & (\sin\beta \cos(\omega-\omega_0) , \sin\beta
\sin(\omega-\omega_0) , \cos\beta ),
\label{eqn:xyzf}
\end{eqnarray}
where $\omega_0$ is the precession angle at which the projection of
$\mathbf{\hat{z}_f}$ onto the BH equatorial plane aligns with the BH
x-axis (see Figure \ref{fig:coord}). This is given by
\begin{equation}
\tan\omega_0 = \frac{\sin i \sin \Phi}{\sin i \cos\Phi \cos\beta -
  \cos i \sin\beta}.
\end{equation}
Note that the flow y-axis always remains in the BH equatorial plane
and the flow x-axis maintains a constant misalignment with the BH
equatorial plane of $\beta$. When $\omega=0^\circ$, the flow is
maximally misaligned ($2\beta$) with the disk, and when $\omega=180^\circ$,
the flow is aligned with the disk. Any point on the surface of the
flow can then be represented in this coordinate system as
\begin{equation}
\mathbf{r} = r ~ ( \sin\theta_f \cos\phi_f , \sin\theta_f \sin\phi_f ,
\cos\theta_f ),
\end{equation}
where $\theta_f$ and $\phi_f$ are respectively the polar and azimuthal
angles defined in the flow coordinate system. The same point can
alternatively be represented in terms of the BH polar and azimuthal
angles
\begin{equation}
\mathbf{r} = r ~ ( \sin\theta \cos\phi , \sin\theta \sin\phi ,
\cos\theta ).
\end{equation}
We present the equations to convert between these two sets of
coordinates in Appendix \ref{sec:BL}. We can also represent a point in
terms of binary polar and azimuthal angles, $\theta_b$ and $\phi_b$.

\subsection{Inner hot flow geometry}
\label{sec:flow_geom}

\begin{figure}
 \includegraphics[height=8.0cm,width=8.5cm,trim=0.0cm 0.0cm 0.0cm
 0.0cm,clip=true]{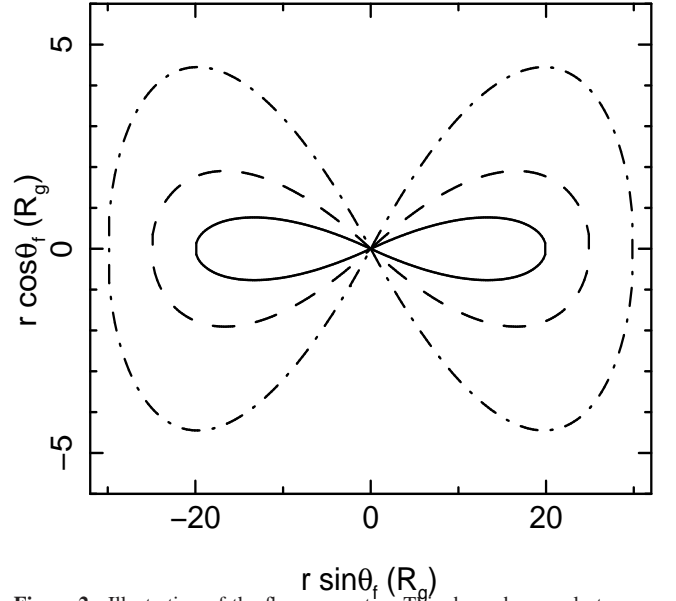}
\vspace{-3mm}
 \caption{Illustration of the flow geometry. This depends on only two
   parameters: the scale height $h/r$ and the outer radius in units of
   $R_g$, $r_o$. For $r \ll r_o$, this function has a constant opening
   angle, but curves around into a tear drop shape  for larger $r$. We
   show three example parameter combinations: $h/r=0.1$, $r_o=20$
   (solid line), $h/r=0.2$, $r_o=25$ (dashed line) and $h/r=0.4$,
   $r_o=30$ (dot-dashed line). In this Paper, we consider the first of
   these geometries, represented by the solid line.}
 \label{fig:flow_geom}
\end{figure}
%Left, Bottom, Right, Top

We assume that the inner flow is shaped like a torus described in
Boyer-Lindquist coordinates by the equation:
\begin{equation}
T(r,\theta_f) = B - \sin\theta_f [ 1 - (1-B)r/r_o ],
\label{eqn:T}
\end{equation}
where $B = [ (h/r)^2 + 1 ]^{-1/2}$, $r_o$ is the outer edge and $h/r$
is the scale height of the flow. Outside of the flow, $T(r,\theta_f) >
0$ and inside the flow, $T(r,\theta_f) < 0 $. On the flow boundary,
$T(r,\theta_f) = 0$. Figure \ref{fig:flow_geom} shows some example
cross-sections of this shape; i.e. contours of $T=0$. For $r \ll r_o$, 
the flow has a constant opening angle, with $\sin\theta_f=B$. For
larger $r$, $\sin\theta_f \geq B$ and the function curves round in the
shape of a tear drop. We also assume an inner radiation edge, $r_{\rm i}$,
which we set equal to the ISCO in this Paper. We can use Equation
(\ref{eqn:T}) to assess if a given position, described by the vector
$\mathbf{r}$, is inside, outside or on the boundary of the flow.

This shape is motivated in part by the discussion on thick disks in
Chapter 10 of \cite{Frank2002}. Equation (\ref{eqn:T}) comes from
considering an inviscid fluid rotating about a central (Newtonian)
point mass (see Figure 10.2 therein). Clearly this is over simplified, 
but it does provide a convenient way to parameterize the flow geometry
in the most realistic way currently possible. Indeed, the shape
illustrated in Figure \ref{fig:flow_geom} is comparable to that seen in
GRMHD simulations (e.g. \citealt{Fragile2007}; \citealt{Fragile2009};
\citealt{Fragile2009a}) which were initialized with a `Polish
doughnut' solution (\citealt{Jaroszynski1980}) and allowed to evolve
self-consistently to show solid body precession. \cite{Dexter2011} used `after the fact'
assumptions about the radiative emissivity of the flow (following
\citealt{Schnittman2006a}) in order to ray trace emission from
the \cite{Fragile2007} simulation but were unable to study long enough
time scales to see the precession period modulate the light curve
(they concentrated instead on trying to find high frequency QPO
candidates). Thus defining a reasonable flow geometry analytically is
currently the best way to study the effects of precession on the
observed emission.

\section{Formalism}
\label{sec:method}

In this Section, we describe our method for calculating the flux and
polarization properties observed from a precessing flow as a function
of precession angle. Since we only consider linear polarization, the
polarization of the signal can be described entirely by the
polarization degree, $p$, and angle, $\chi$.

\subsection{Inner hot flow properties}

\begin{figure}
 \includegraphics[height=8.0cm,width=8.5cm,trim=0.0cm 0.0cm 0.0cm
 0.0cm,clip=true]{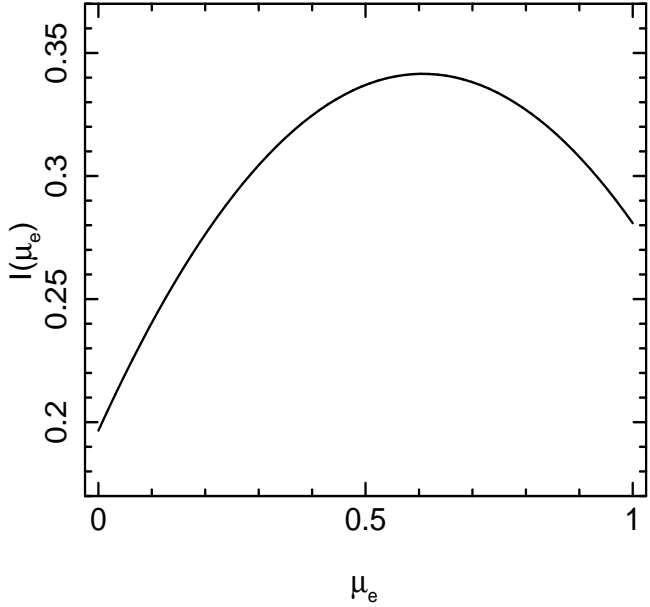}
\vspace{0mm}
 \caption{Intensity as a function of emission angle assumed for this
   Paper. We use the analytic function $I(\mu_e) \propto 1 +
   1.1(1-\mu_e) - 1.4(1-\mu_e)^2$, designed to mimic the results of
   ST85 for the $\tau=1$ case. This Figure can be compared
   with Figure 4 of ST85, except we use a different normalization.}
\vspace{5mm}
 \label{fig:intmu}
\end{figure}
%Left, Bottom, Right, Top

\begin{figure}
 \includegraphics[height=8.0cm,width=8.5cm,trim=0.0cm 0.0cm 0.0cm
 0.0cm,clip=true]{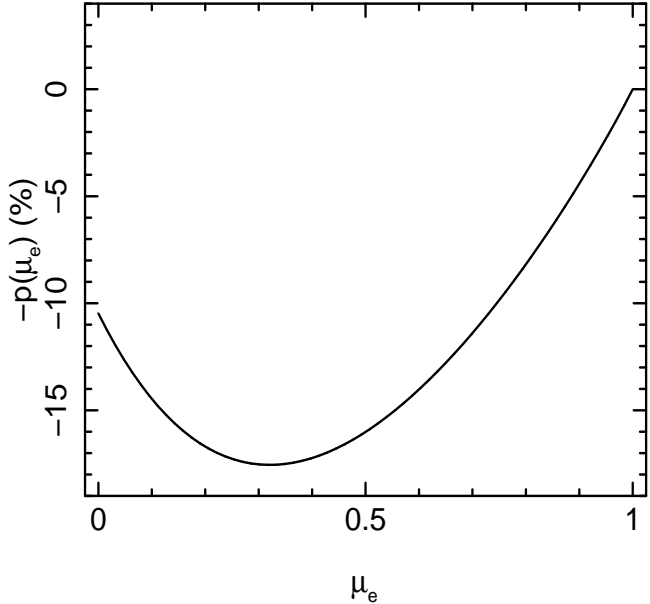}
\vspace{0mm}
 \caption{Polarization degree as a function of emission angle assumed
   for this Paper. We use the analytic function $p(\mu_e) =
   50\%(1-\mu_e) - 23\%[\exp{ \{ (1-\mu_e)^{1.6} \} }-1]$, designed to mimic
   the results of ST85 for the $\tau_0=1$ case. This Figure can be compared
   with Figure 5 of ST85.}
\vspace{6mm}
 \label{fig:polmu}
\end{figure}
%Left, Bottom, Right, Top

The specific intensity emitted from a patch of the flow surface is, in
general, a function of position, photon energy and emission angle,
$\theta_e$. Our calculations are greatly simplified by assuming that
these variables can be separated, such that
\begin{equation}
I_{E_e}(E_e,r,\mu_e) = I_{E_e} ~ q_e(r) ~ I(\mu_e),
\label{eqn:sep}
\end{equation} 
where $I_{E_e}$ is the emitted spectrum, $q_e(r)$ and $I(\mu_e)$ are
respectively the radial and angular emissivity profiles and
$\mu_e=\cos\theta_e$. In the absence of a standard radial emissivity
law for a large scale height accretion flow, we simply assume the
standard case for a thin disk
(\citealt{Shakura1973}; \citealt{Novikov1973})
\begin{equation}
q_e(r) = r^{-3} ( 1 - \sqrt{r_{\rm i} / r }),
\end{equation}
following VPI13. As for the angular profile, we define an analytic
function designed to mimic the shape obtained from the calculations
of ST85. We plot this profile in Figure
\ref{fig:intmu}, to be compared with the $\tau_0=1$ case from Figure 4
of ST85. Here, we have normalised the intensity to give
\begin{equation}
2\pi \int_0^1 I(\mu_e) \mu_e {\rm d} \mu_e = 1.
\end{equation}

We also parameterize polarization degree as a function of emission
angle $p(\mu_e)$ (defined as the fraction of photons emerging from
the flow which are polarized) following the calculations of
ST85. Figure \ref{fig:polmu} shows this function, to be compared with
the $\tau_0=1$ case in Figure 5 of ST85. We plot $-p(\mu_e)$ in order
to account for the different sign convention in ST85.
We note that the emission angle $\theta_e$ is defined here
  with respect to the flow spin axis rather than to the local normal
  to the torus surface. This is because the calculations of ST85 (also
  see \citealt{Viironen2004}) assume Thomson scattering from a thin
  slab and the functions in Figures (\ref{fig:intmu}) and
  (\ref{fig:polmu}) are therefore only defined with respect to the
  flow spin axis. Since we expect the flow to have a rather small
  scale height ($h/r\sim 0.1$) this is likely a fairly good
  assumption. There is however scope to improve upon these assumptions
  in future work. For instance, \cite{Poutanen1996} considered cases
  where exact solutions can be found outside of the Thomson scattering
  limit.

\subsection{Ray tracing}

We use the publicly available code \textsc{geokerr}, described in
\cite{Dexter2009}, to solve for photon geodesics in the Kerr
metric. We start off by defining an observer's camera some large
distance, $D$, from the BH\footnote{The camera must be
  far enough away for all geodesics to be straight and parallel. We
  use $D=10^5~R_g$.} along the vector $\mathbf{\hat{o}}$. The
impact parameters at infinity, $\alpha_0$ and $\beta_0$, represent
respectively horizontal and vertical distance on the plane of the
observer's camera (a schematic illustrating the impact parameters can
be found to the left of Figure A1 in \citealt{Middleton2015}). For a
given observer position and BH spin, $a$, null geodesics can be
uniquely defined by these two impact parameters. Alternatively, these
geodesics can be parameterised by Carter's constants of motion $l =
-\alpha_0 \sin\theta_0$ and $q^2 = \beta_0^2 + \cos^2\theta_0 (
\alpha_0^2 - a^2)$. Each combination of impact parameters represents a
pixel on the observer's camera, which is hit by a photon on a unique
geodesic path. We define a grid of impact parameters with equal
logarithmic steps in $b=\sqrt{\alpha_0^2 + \beta_0^2}$ and equal
linear steps in the angle $\varphi$, defined as $\tan\varphi =
\alpha_0/\beta_0$. Ignoring parallax, which is the same for each
pixel, the solid angle subtended by each pixel is $b~{\rm d}b {\rm
  d}\varphi$.

For each pixel we set \textsc{geokerr} to compute 100 steps along the
photon geodesic towards the BH. For each step, a position 4-vector
$x^\mu = (t,r,\theta,\phi)$ is provided in Boyer-Lindquist
coordinates. We convert the BH polar and azimuthal angles $\theta$ and
$\phi$ to the corresponding flow angles $\theta_f$ and $\phi_f$ using
Equations (\ref{eqn:muf}) and (\ref{eqn:tanf}). This allows us to
evaluate the function $T(r,\theta_f)$ from Equation (\ref{eqn:T}). If
$T$ passes from positive to negative between consecutive steps, we
conclude that the geodesic has crossed the flow boundary. If this does
not happen, we also check if $\mu_f \equiv \cos\theta_f$ has passed
from positive to negative, in which case the geodesic has crossed the
flow mid-plane and therefore must have crossed into \textit{and} out of
the flow in a single step. This is most likely to happen for small $r$
where the flow is very shallow and a step does not need to be
particularly large to pass completely through the flow.

In either of these cases, we have isolated a root of Equation
(\ref{eqn:T}) between two points on the geodesic path. We use linear
interpolation to represent $r$ as a function of $\mu_f$ between these
two points, allowing us to express $T$ as a function of $\mu_f$
only. We then find the root of the equation $T(\mu_f)=0$ with a
bisection search. From this solution for $\mu_f$, we can interpolate a
solution for $r$ and also $\phi_f$. If $r<r_{\rm i}$, we carry on
following the geodesic in case it loops back around to intercept the
underside of the flow. Otherwise, we calculate the contribution to the
flux observed from that pixel. We ignore direct disk emission and also
emission from the flow reflected from the outer disk. This assumption
is appropriate for the energy range $\sim 10-20$ keV, which is
dominated by Comptonized emission from the flow. Direct disk emission
contributes only in soft X-rays $\lesssim 5$ keV and reflection is
important at the iron line ($\sim 6.4$ keV) and above $\sim 20$ keV.

\subsection{Disk shielding}

When tracing rays backwards from the observer, we also test whether
they intercept the outer disk before hitting the flow. In this case,
since the disk is optically thick, we conclude that our view of the
flow is blocked for this pixel. We assume that the outer disk occupies
the binary plane and truncates at $r=r_{\rm o}$. At every step along
the ray, we convert to binary coordinates to assess if $\mu_b\equiv
\cos\theta_b$ passes from positive to negative, indicating a crossing
of the disk plane. If this happens, we use linear interpolation to
calculate the value of $r$ at $\mu_b=0$. If $r>r_{\rm o}$, we conclude
the ray has crossed the outer disk and stop following the
ray. Otherwise, we carry on following it. In the cases where the ray
crosses both the disk and flow in one step, we use linear
interpolation to assess which it hit first.

%ps2eps -B -g justbs01_i70_F110_b10.ps
\begin{figure*}
\centering
\includegraphics[height=7cm,width=6.5cm,trim=0.0cm 0.0cm 0.0cm
 0.0cm,clip=true]{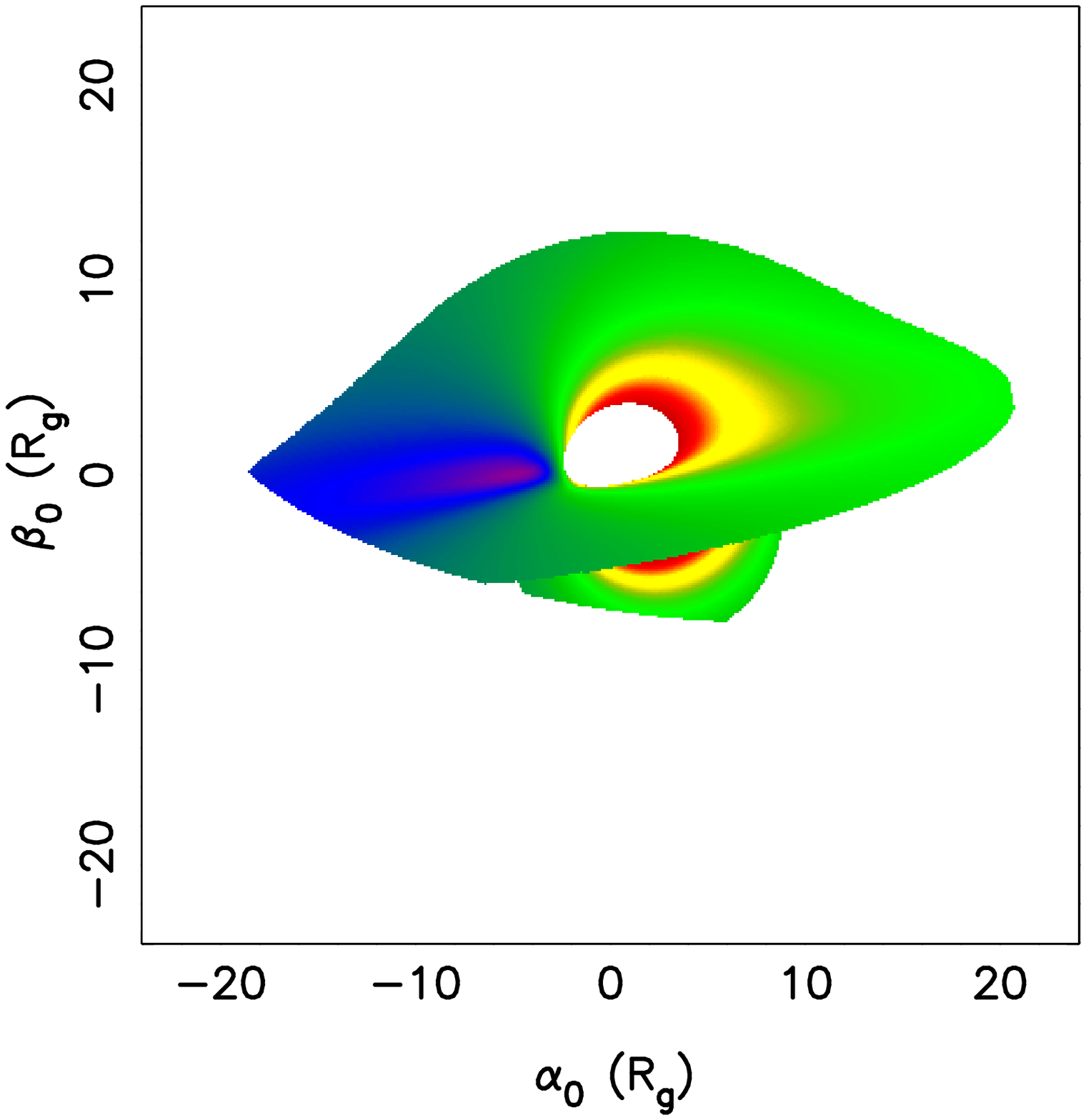}
\includegraphics[height=7cm,width=6.5cm,trim=0.0cm 0.0cm 0.0cm
 0.0cm,clip=true]{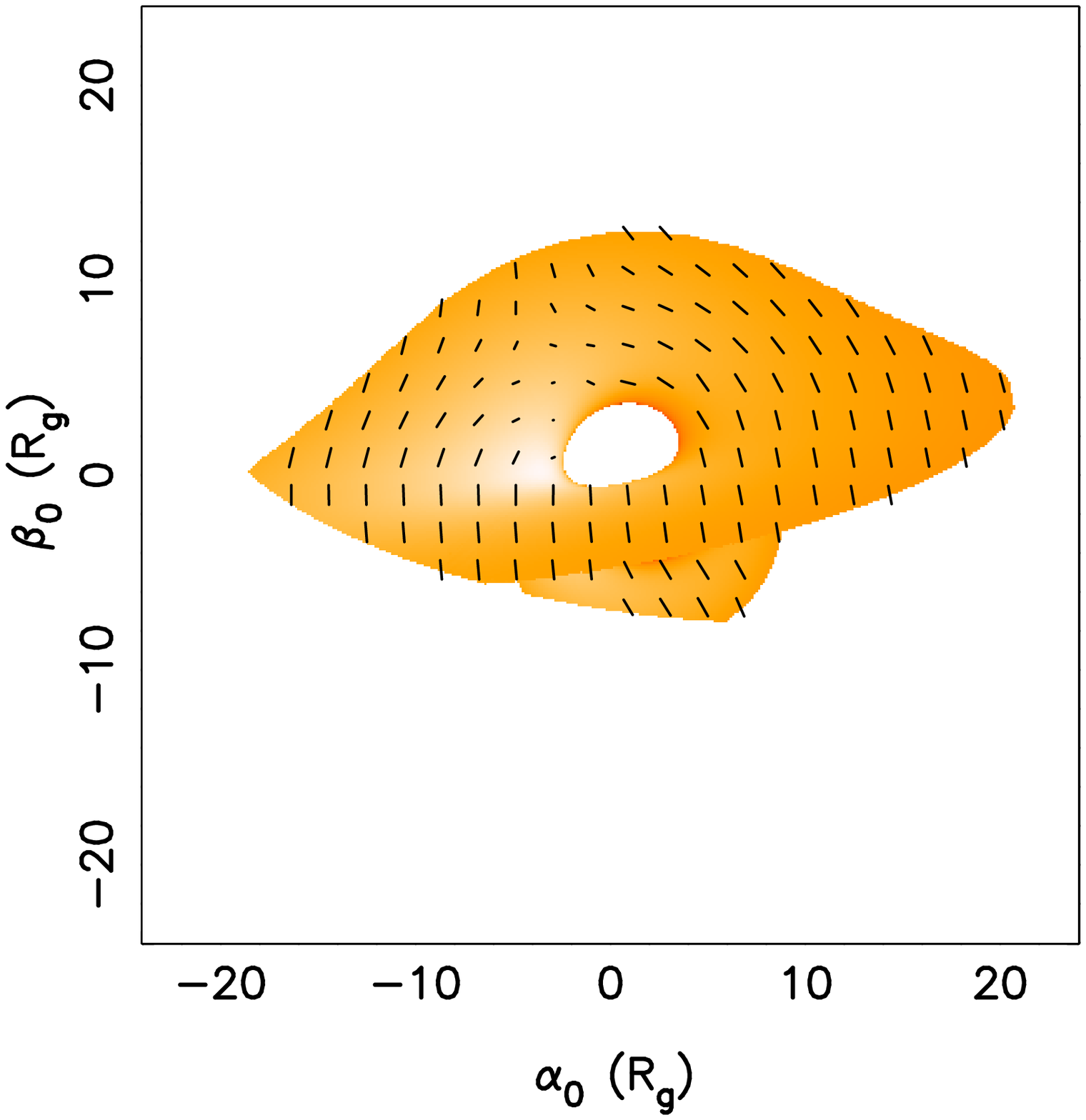} \\
\includegraphics[height=7cm,width=6.5cm,trim=0.0cm 0.0cm 0.0cm
 0.0cm,clip=true]{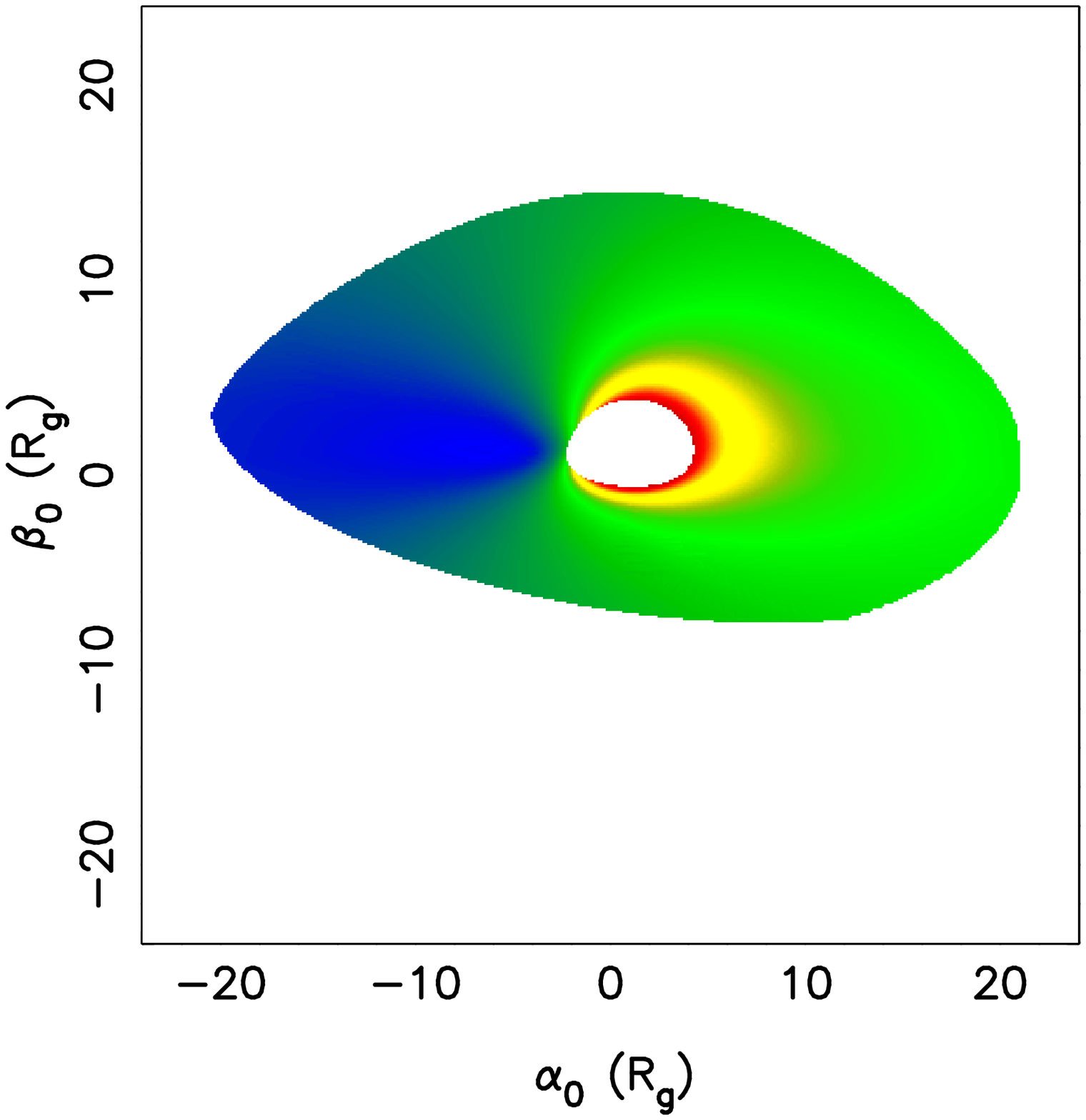}
\includegraphics[height=7cm,width=6.5cm,trim=0.0cm 0.0cm 0.0cm
 0.0cm,clip=true]{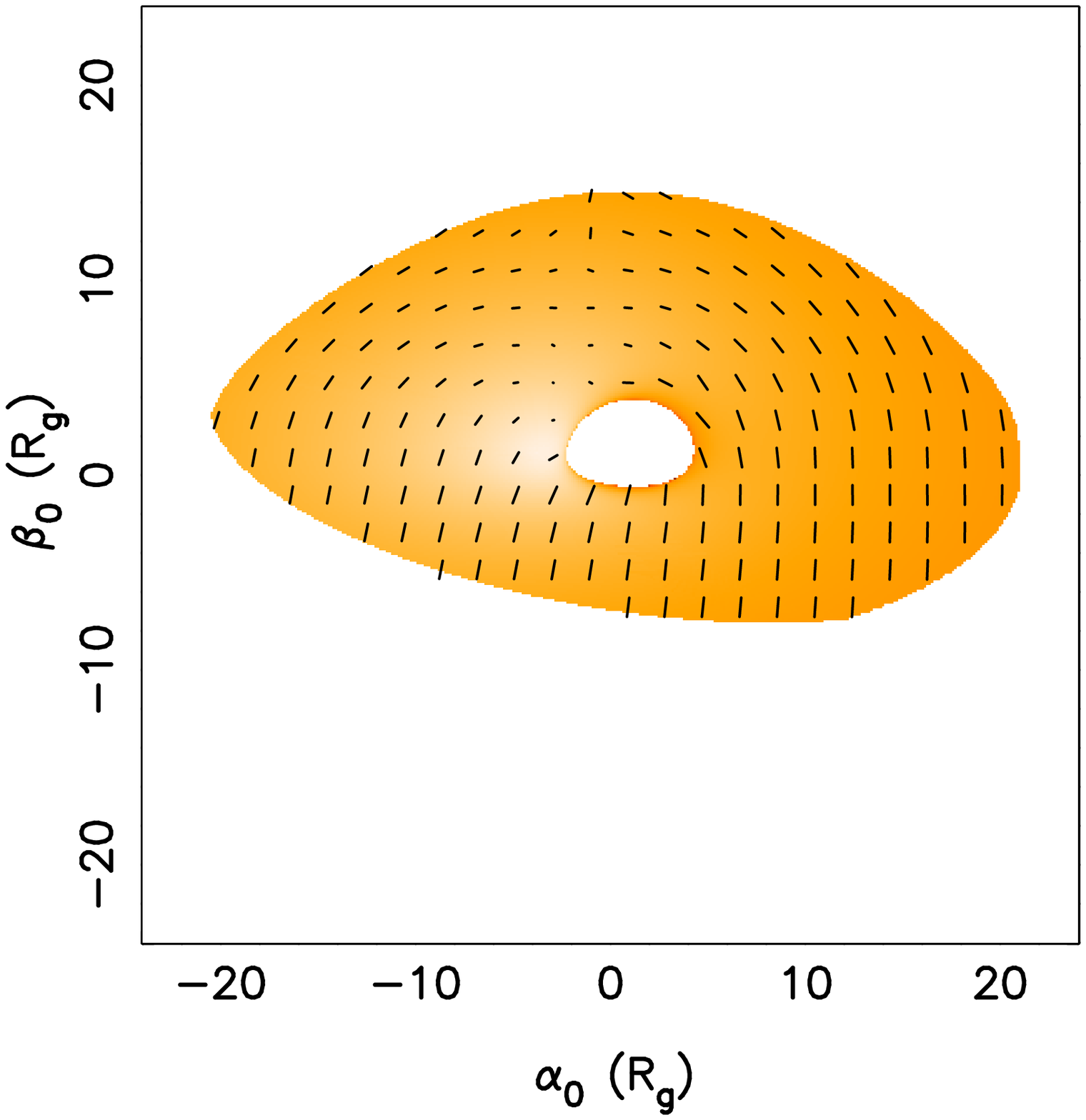} \\
\includegraphics[height=7cm,width=6.5cm,trim=0.0cm 0.0cm 0.0cm
 0.0cm,clip=true]{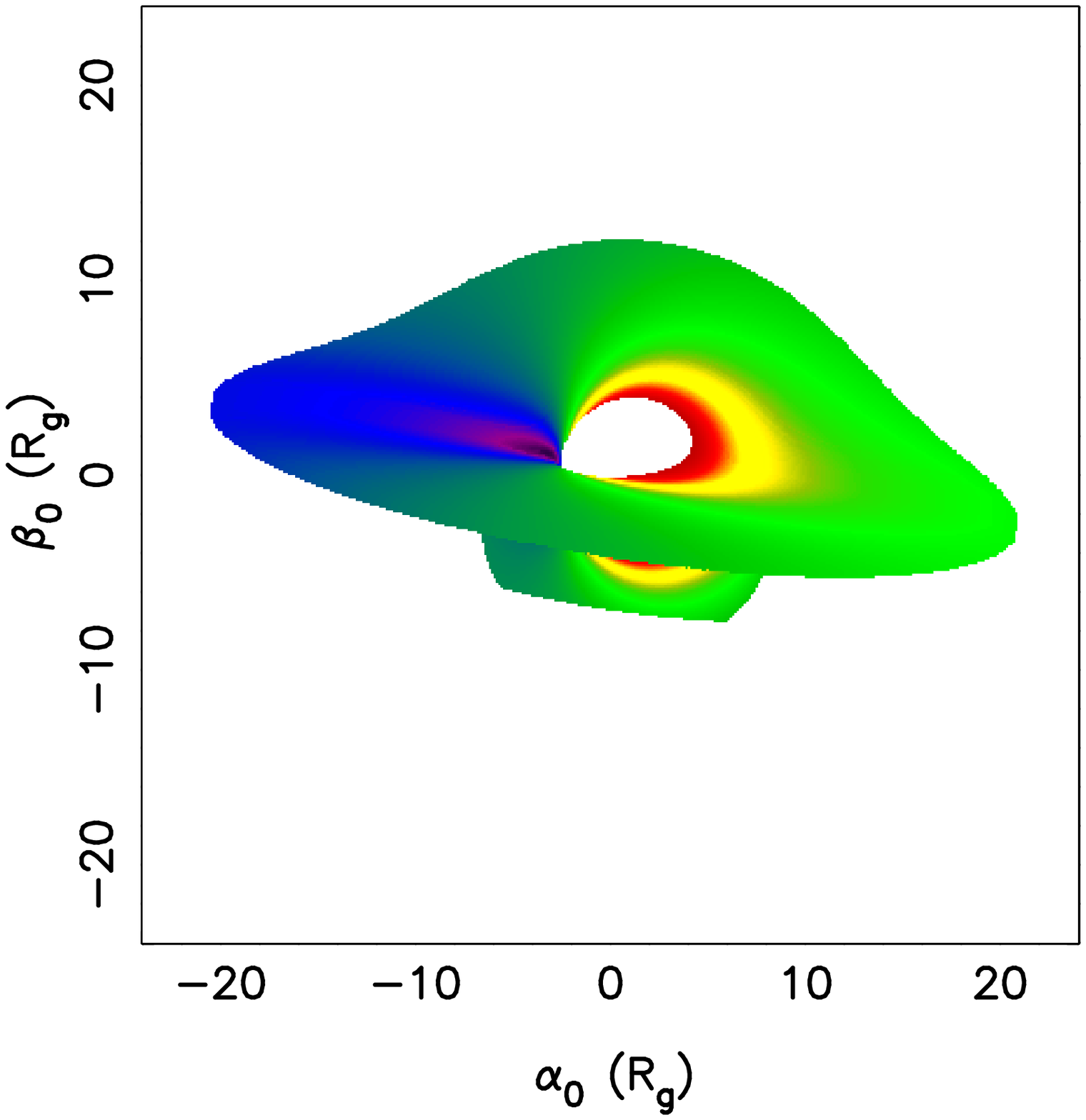}
\includegraphics[height=7cm,width=6.5cm,trim=0.0cm 0.0cm 0.0cm
 0.0cm,clip=true]{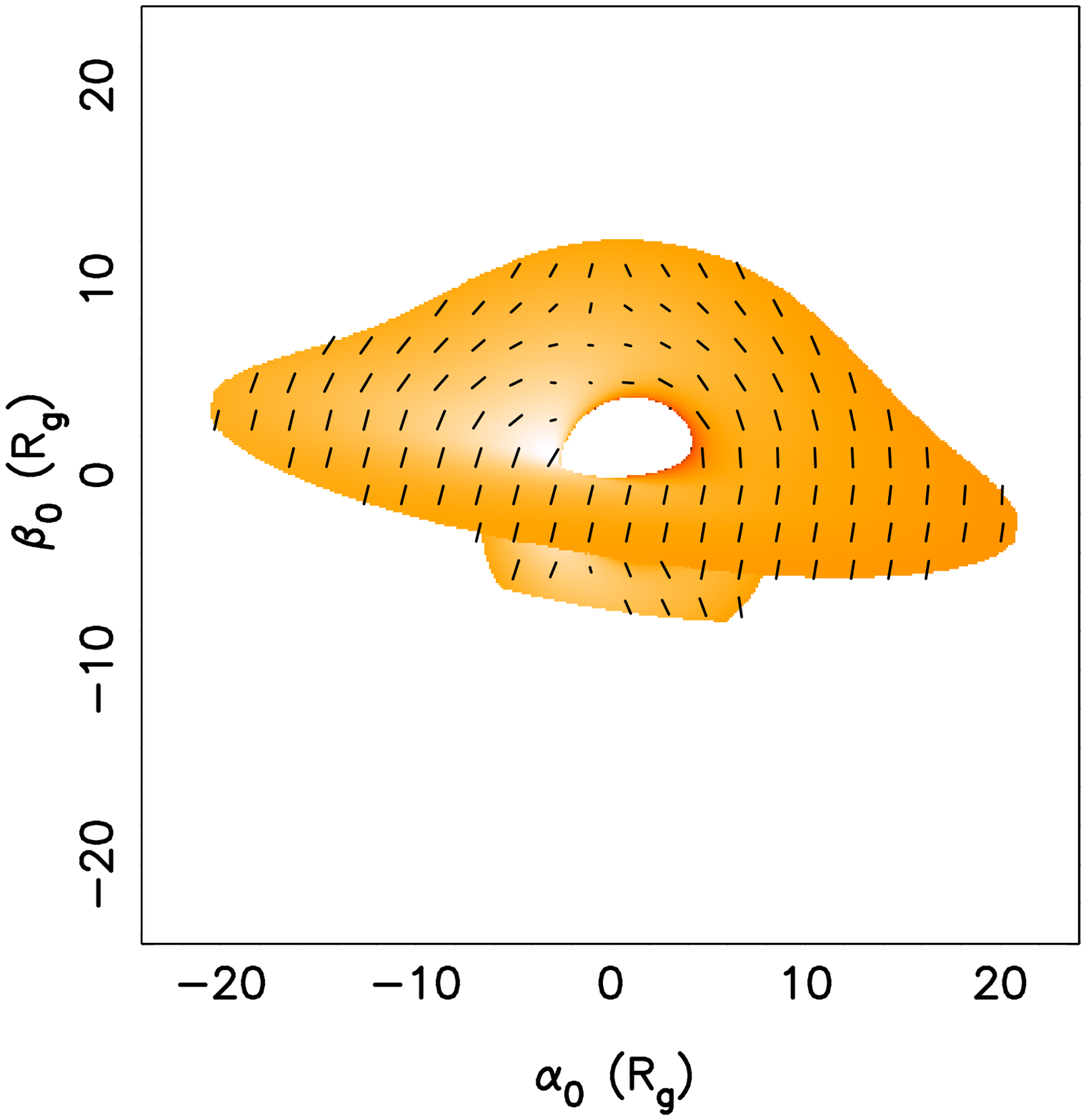} \\
~~~~~~~\includegraphics[height=5.5cm,width=1.5cm,trim=-0.3cm 0.6cm 0.0cm
 0.0cm,clip=true, angle=-90]{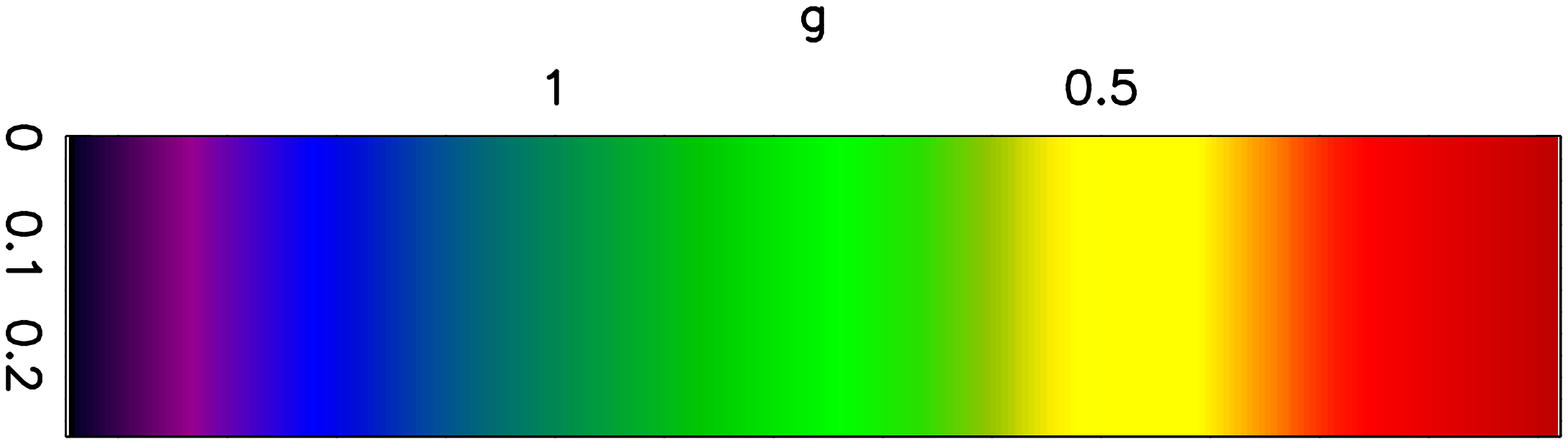} ~~~~~~~~~~~~~~~~
\includegraphics[height=5.5cm,width=1.5cm,trim=0.0cm 0.6cm 0.0cm
 0.0cm,clip=true, angle=-90]{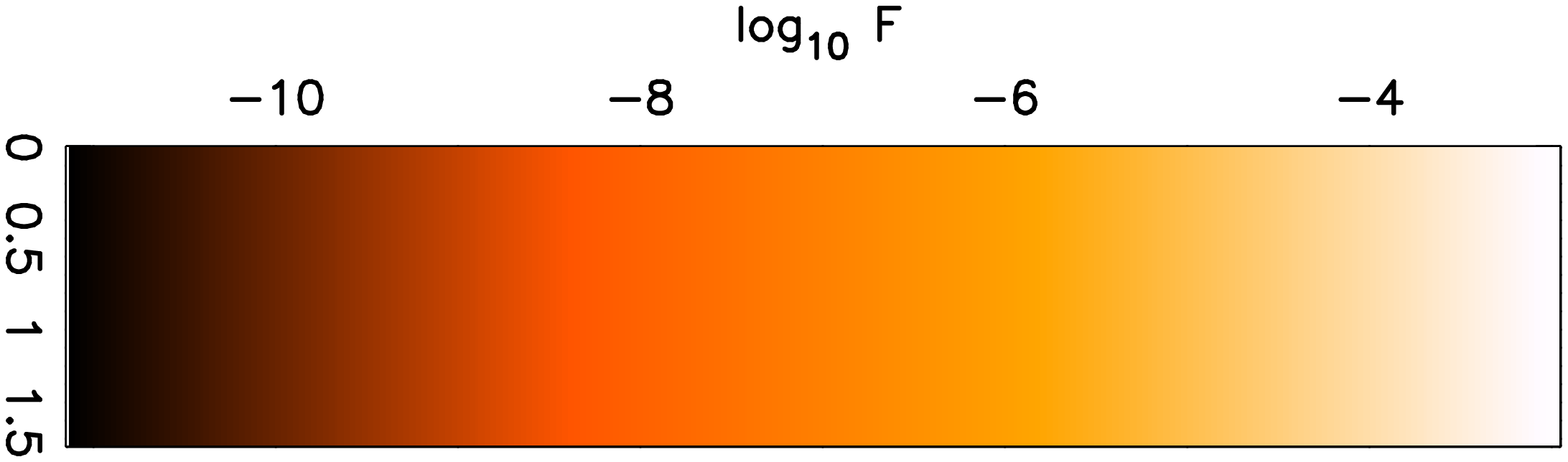}
\caption{Images of the flow for three precession angles for the model
   with $i=70^\circ$, $\Phi=110^\circ$, $\beta=10^\circ$ and
   $h/r=0.1$. The left hand side pictures blueshift and the right hand
   side pictures flux with the polarization vector overlaid,
   normalised to the maximum observed polarization degree. The three
   precession angles pictured, in units of cycles, are $\omega=0$,
   $0.3125$ and $0.625$ from top to bottom respectively. The full
   movies for these images can be found at {\footnotesize \color{blue}
  \underline{\smash{http://figshare.com/articles/Polarization\_modulation\_gifs/1351920}}
  \color{black}}. Indivdual movies can alternatively be found on YouTube: {\footnotesize \color{blue}
  \underline{\smash{www.youtube.com/watch?v=Q2CwOGKVC9U\&feature=youtu.be}}
  \color{black}} (blueshift) and {\footnotesize
  \color{blue}\underline{\smash{www.youtube.com/watch?v=E3kYAnS3pQI\&feature=youtu.be}}  
  \color{black}} (flux and polarization).}
\label{fig:i70an}
\end{figure*}

\subsection{Blueshift and flux}

The energy of a photon reaching the observer will be modified by the
gravitational field of the BH and the motion of the emitting
particle. For a stationary observer at infinity, the ratio of the
observed to emitted energy of a photon (hereafter the blueshift) is given by
\begin{equation}
\mathtt{g} \equiv \frac{E_o}{E_e} = \frac{-p_o^t}{p_e^\mu u_{\mu}},
\label{eqn:gfac}
\end{equation}
where $p_e^\mu$ and $u^\mu$ are respectively the 4-momentum of the
photon and the 4-velocity of the emitting particle
(e.g. \citealt{Luminet1979}). We use a form for the 4-momentum (given
in full in the Appendix) normalised such that $p_o^t = 1$,
simplifying Equation (\ref{eqn:gfac}) further. Since we use the
metric convention of negative time entries and positive distance
entries, the 4-velocity of the emitting particle is $(u^\mu)^2 =
-(dx^\mu/ds)^2$, where $ds^2 = g_{\mu\nu}~dx^\mu~dx^\nu$ is the line 
element and $g_{\mu\nu}$ is the metric. We can represent this in terms
of coordinate velocities, $\Omega^\mu \equiv dx^\mu / dt$, in the form
\begin{equation}
u^\sigma = \frac{\Omega^\sigma}{ \sqrt{-g_{\mu\nu}
    \Omega^\mu\Omega^\nu }}.
\label{eqn:fourvel}
\end{equation}
We assume circular orbits ($\Omega^r=0$) with Keplerian angular
velocity about the flow spin axis, so
\begin{equation}
\Omega^{\phi_f} \equiv \frac{d\phi_f}{dt} = \frac{1}{r^{3/2}+a}
\end{equation}
and $\Omega^{\theta_f}=0$. Note, we assume Keplerian angular
  velocity perpendicular to the flow spin axis even for parts of the
  flow not in the mid-plane. Since $\Omega^{\phi_f} \gg \Omega^\omega =
2\pi \nu_{qpo}$, we convert this to BH coordinates as follows:
\begin{eqnarray}
\Omega^{\phi} &=& \frac{\partial \phi}{\partial \phi_f}
\Omega^{\phi_f} \nonumber \\
\Omega^{\theta} &=& \frac{\partial \theta}{\partial \phi_f}
\Omega^{\phi_f},
\end{eqnarray}
where the differentials can be computed from Equations (\ref{eqn:mu})
and (\ref{eqn:tan}). We use the Kerr metric throughout, which we
provide in Appendix \ref{sec:kerr} for completeness.

The specific flux observed at energy $E_o$ from a pixel with solid
angle $b~{\rm d}b ~{\rm d}\varphi$ is given by (\citealt{Luminet1979})
\begin{equation}
dF_{E_o} = \mathtt{g}^3 I_{E_e}(E_e,r,\mu_e) b~{\rm d}b ~{\rm d}\varphi.
\label{eqn:flux1}
\end{equation}
For energies significantly greater than the seed photon temperature
($kT_{\rm bb}$) and less than the electron temperature ($kT_{\rm e}$),
the emitted spectrum can be well approximated by a power law, $I_{E_e}
\propto E_e^{1-\Gamma}$. In this case, the observed spectrum is also a
power law, $I_{E_o} \propto E_o^{1-\Gamma}$. Substituting this (and
Equation \ref{eqn:sep}) into Equation (\ref{eqn:flux1}), we can
represent the flux in a power law dominated energy band ($\sim 10-20$
keV) as
\begin{equation}
dF = \mathtt{g}^{2+\Gamma} I_{E_o} q_e(r) I(\mu_e) b~{\rm d}b~ {\rm d}\varphi.
\label{eqn:flux2}
\end{equation}

The emission angle is given by (e.g. \citealt{Misner1973};
\citealt{Rybicki1997}; \citealt{Dovciak2004})
\begin{equation}
\mu_e = \mathtt{g}~p_e^\mu n_\mu,
\end{equation}
where $n^\mu$ is a 4-vector normal to the flow mid-plane, defined in
the flow rest-frame. See Appendix \ref{sec:tetrad} for the full form
of this 4-vector. The above formula encapsulates the effects of both
light bending and relativistic aberration. From this, we evaluate $dF$
for every pixel and sum to obtain the total observed flux, $F$, as a
function of precession angle, $\omega$.

\subsection{Polarization}
\label{sec:polar} 

For each geodesic that intercepts the flow, we calculate the
polarization degree from the function shown in Figure
\ref{fig:polmu}. We initialize the polarization 4-vector, $f^\mu$, as
the projection of $n^\mu$ on the plane perpendicular to the emergent
photon's 4-momentum, $p^\mu$. Following, \cite{Dovciak2004}, this is
defined as
\begin{equation}
f^\mu = \frac{ n^\mu - \mu_e ( \mathtt{g} p_e^\mu - u^\mu ) }
{ \sqrt{1-\mu_e^2} }.
\end{equation}
For each geodesic, we parallel transport the polarization 4-vector
forwards from the emission point to the observer's camera in geodesic
steps, $\delta x_\mu$. That is, moving from $s$ to $s+\delta s$
changes the polarization 4-vector to
\begin{equation}
f^\mu(s+\delta s) = f^\mu(s) - f^\sigma(s) \Gamma^\mu_{\sigma\nu} \delta x^\nu,
\end{equation}
with the Christoffel symbols given by
\begin{equation}
\Gamma^\sigma_{\mu\nu} = \frac{1}{2} g^{\sigma\kappa} \left(
   \frac{ \partial g_{\nu\kappa} }{ \partial x^\mu }
+\frac{ \partial g_{\kappa\mu} }{ \partial x^\nu }
- \frac{ \partial g_{\mu\nu} }{ \partial x^\kappa } \right).
\end{equation}
The position 4-vector, $x^\mu$, is specified for each point on each
geodesic by \textsc{geokerr} and so we can simply calculate the step
length as $\delta x^\mu = x^\mu(s+\delta s) - x^\mu(s)$.

Once we have transported $f^\mu$ all the way back to the observer, we
convert back to BH Cartesian coordinates (see Appendix \ref{sec:BL})
and find the projection of $f^\mu$ on the horizontal and vertical
directions of the camera. These directions are given in BH (Cartesian)
coordinates by
\begin{eqnarray}
\mathbf{\hat{\alpha}_0} &=& ( 0 , 1 , 0 ) \nonumber \\
\mathbf{\hat{\beta}_0} &=& ( -\cos\theta_0 , 0 , \sin\theta_0 ),
\end{eqnarray}
and the projections are simply $\mathbf{\hat{f}} \cdot
\mathbf{\hat{\alpha}_0}$ and $\mathbf{\hat{f}} \cdot
\mathbf{\hat{\beta}_0}$, where $\mathbf{\hat{f}}$ is the 3-vector
describing the spatial part of $f^\mu$. The polarization angle for a
given geodesic is then given by
\begin{equation}
\tan[ \chi(\alpha_o,\beta_0) ] = -\frac{ \mathbf{\hat{f}} \cdot \mathbf{\hat{\alpha}_0} }  {
  \mathbf{\hat{f}} \cdot \mathbf{\hat{\beta}_0} }.
\end{equation}
We define Stokes parameters for each geodesic
\begin{eqnarray}
dQ & = & dF(\alpha_o,\beta_0)~p(\alpha_o,\beta_0)~\cos[
2\chi(\alpha_o,\beta_0) ] \\
dU & = & dF(\alpha_o,\beta_0)~p(\alpha_o,\beta_0)~\sin[
2\chi(\alpha_o,\beta_0) ],
\end{eqnarray}
and sum over all geodesic paths to get the observed Stokes
parameters $Q=\int dQ$ and $U = \int dU$. Note that, under this
convention, vertical and horizontal polarization corresponds to
$Q/F=1$ and $Q/F=-1$ respectively for a $100\%$ polarized signal. The
overall polarization degree is then
\begin{equation}
p = \frac{ \sqrt{Q^2+U^2} }{F}
\end{equation}
and the overall polarization angle, $\chi$, can be calculated from
\begin{equation}
\tan( 2\chi ) = \frac{ U }{ Q }.
\end{equation}

\subsection{Simple Newtonian approximation}
\label{sec:newt}

In addition to the fully relativistic treatment described above, we
also define a simple Newtonian approximation to assess the importance
of GR effects. We make the simplifying assumption that the flow is a
flat disk, and so any point on the flow surface is described by the
vector $\mathbf{r} = x\mathbf{\hat{x}_f}+y\mathbf{\hat{y}_f}$. In a
flat metric, photons emitted from $\mathbf{r}$ hit the observer's
camera at the coordinates $\alpha_0=\mathbf{r}\cdot
\mathbf{\hat{\alpha}_0}$ and $\beta_0=\mathbf{r}\cdot
\mathbf{\hat{\beta}_0}$. The Keplerian velocity (in units of $c$) has
magnitude $v=r^{-1/2}$ and direction $\mathbf{\hat{v}} = (-y
\mathbf{\hat{x}_f} + x \mathbf{\hat{y}_f}) / \sqrt{x^2+y^2}$. We use
Equation (\ref{eqn:flux2}) to calculate the flux for each pixel of the
camera. Using the Minkowski metric instead of the Kerr metric for
Equations (\ref{eqn:fourvel}) and (\ref{eqn:gfac}), the blue shift
becomes
\begin{equation}
\mathtt{g} = \frac{ \sqrt{1-v^2} } {1- \mathbf{\hat{o}}\cdot \mathbf{v} }.
\end{equation}
This is the special relativistic Doppler factor. Since all the photons
emitted at a given time now have the same polarization angle, the
observed polarization vector $\mathbf{\hat{f}}$ can be calculated as the
projection of $\mathbf{\hat{z}_f}$ onto the plane perpendicular to
$\mathbf{\hat{o}}$ (if relativistic aberration is ignored). This gives
for the polarization angle
\begin{equation}
\tan\chi = -\frac{\sin\beta \sin(\omega-\omega_0)}
{\sin\theta_0 \cos\beta - \cos\theta_0 \sin\beta \cos(\omega-\omega_0)},
\label{eqn:newtchi}
\end{equation}
which is the standard formula for a rotating vector
(\citealt{Ferguson1973}; \citealt{Ferguson1976};
\citealt{Viironen2004}). Moreover, the observed polarization degree
simply becomes a flux weighted average of the contribution from each
pixel.

\section{Results}
\label{sec:results}

%ps2eps -B -g -f Stokes4_i70_F110_b10.ps
\begin{figure}
 \includegraphics[height=12.0cm,width=8.0cm,trim=0.0cm 0.0cm 0.0cm
 0.0cm,clip=true]{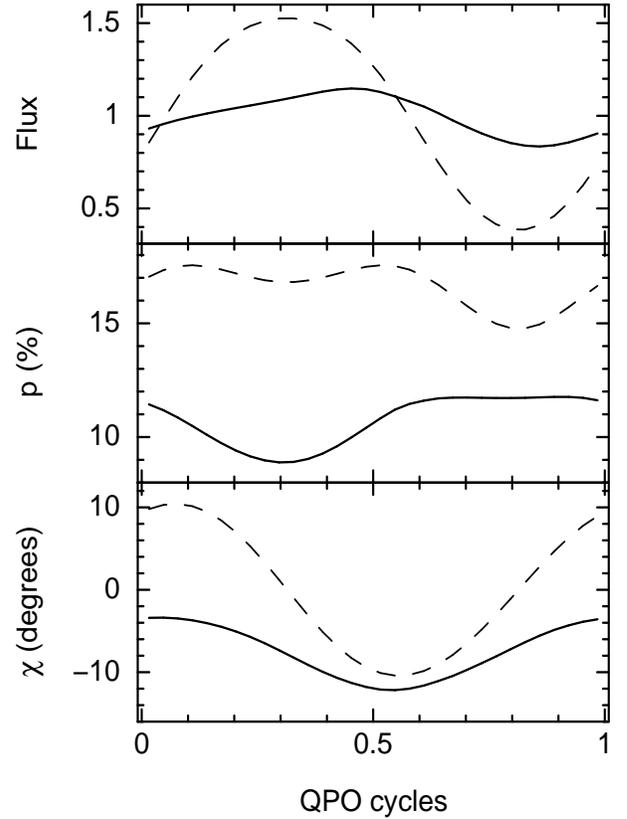}
\vspace{0mm}
 \caption{Flux, polarization degree and polarization angle plotted
   against QPO phase for the high inclination model. The solid lines
   are for the fully relativistic model and dashed lines representing
   a simple Newtonian approximation are included for comparison. The
   fractional rms (for the full model) in the first and second
   harmonics respectively is $10.0\%$ and $2.24\%$ for the flux. The
   \textit{absolute} rms in the first and second harmonics is
   respectively $1.0\%$ and $0.3\%$ for the polarization degree and
   respectively $3.1^\circ$ and $0.2^\circ$ for the polarization angle.}
 \label{fig:i70stoke}
\end{figure}
%Left, Bottom, Right, Top

%ps2eps -B -g -f Stokes4_i30_F180_b10.ps
\begin{figure}
 \includegraphics[height=12.0cm,width=8.0cm,trim=0.0cm 0.0cm 0.0cm
 0.0cm,clip=true]{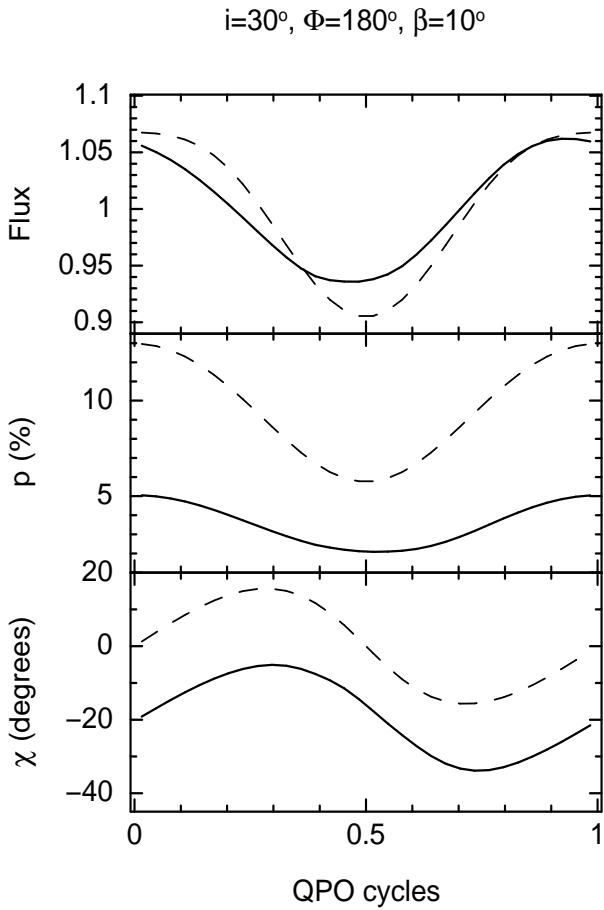}
\vspace{0mm}
 \caption{Flux, polarization degree and polarization angle plotted
   against QPO phase for the low inclination model. Solid and dashed
   lines again refer to the fully relativistic model and the Newtonian
   comparison respectively.  The
   fractional rms (for the full model) in the first and second
   harmonics respectively is $4.6\%$ and $0.25\%$ for the flux. The
   \textit{absolute} rms in the first and second harmonics is
   respectively $1.1\%$ and $0.05\%$ for the polarization degree and
   respectively $10.0^\circ$ and $1.0^\circ$ for the polarization angle.}
 \label{fig:i30stoke}
\end{figure}
%Left, Bottom, Right, Top

\begin{figure*}
\centering
\includegraphics[height=7cm,width=6.5cm,trim=0.0cm 0.0cm 0.0cm
 0.0cm,clip=true]{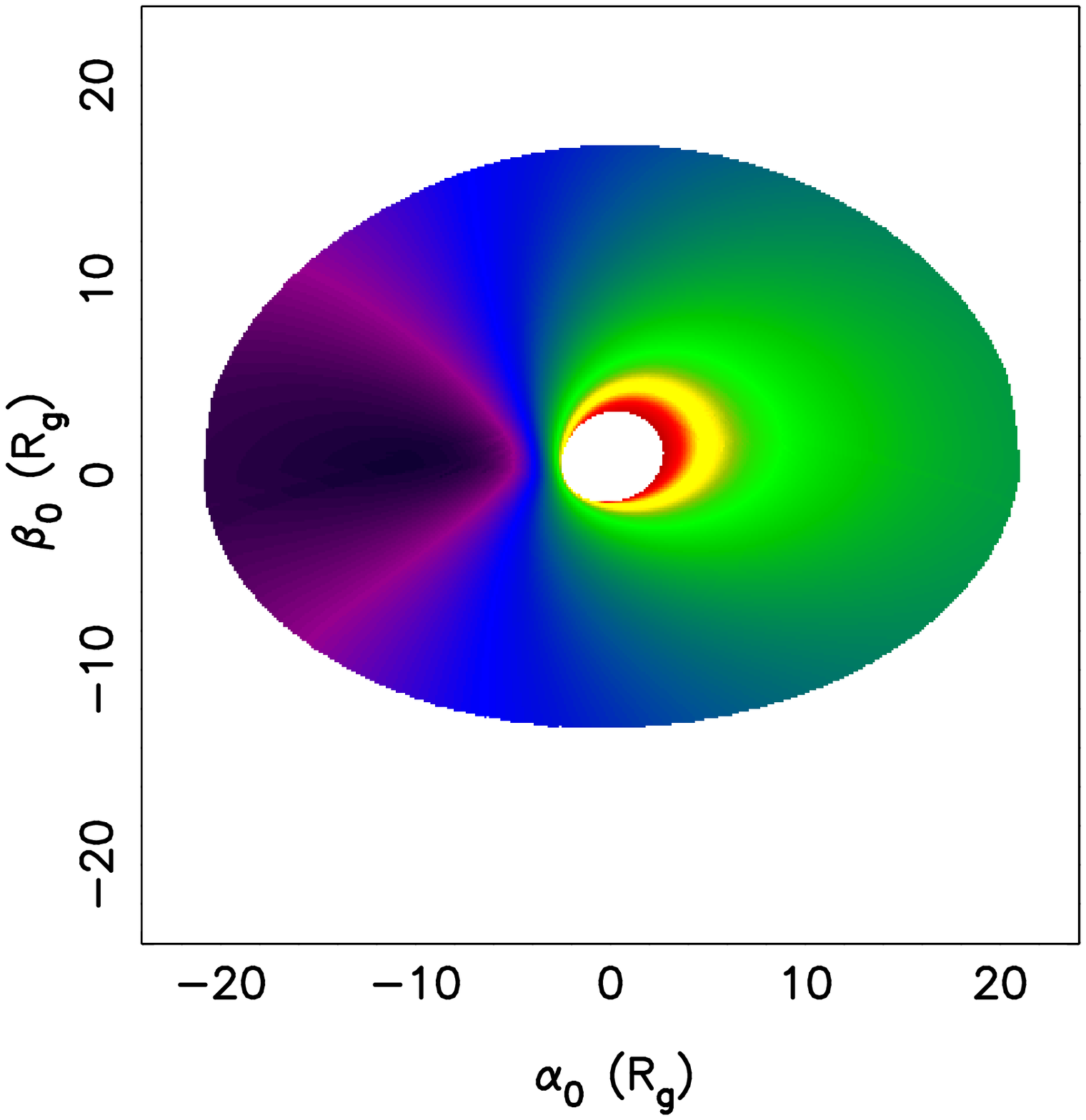}
\includegraphics[height=7cm,width=6.5cm,trim=0.0cm 0.0cm 0.0cm
 0.0cm,clip=true]{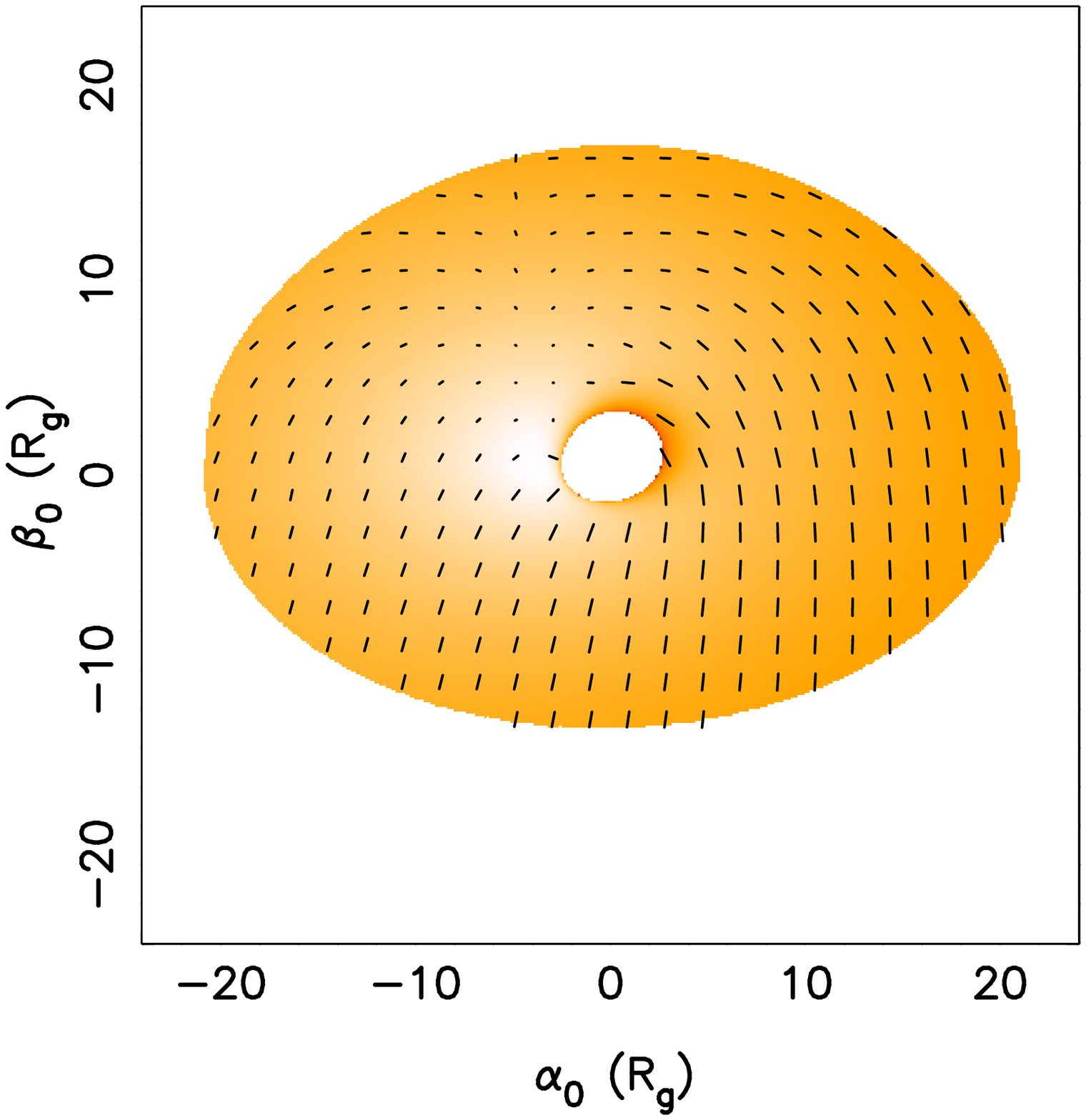} \\
\includegraphics[height=7cm,width=6.5cm,trim=0.0cm 0.0cm 0.0cm
 0.0cm,clip=true]{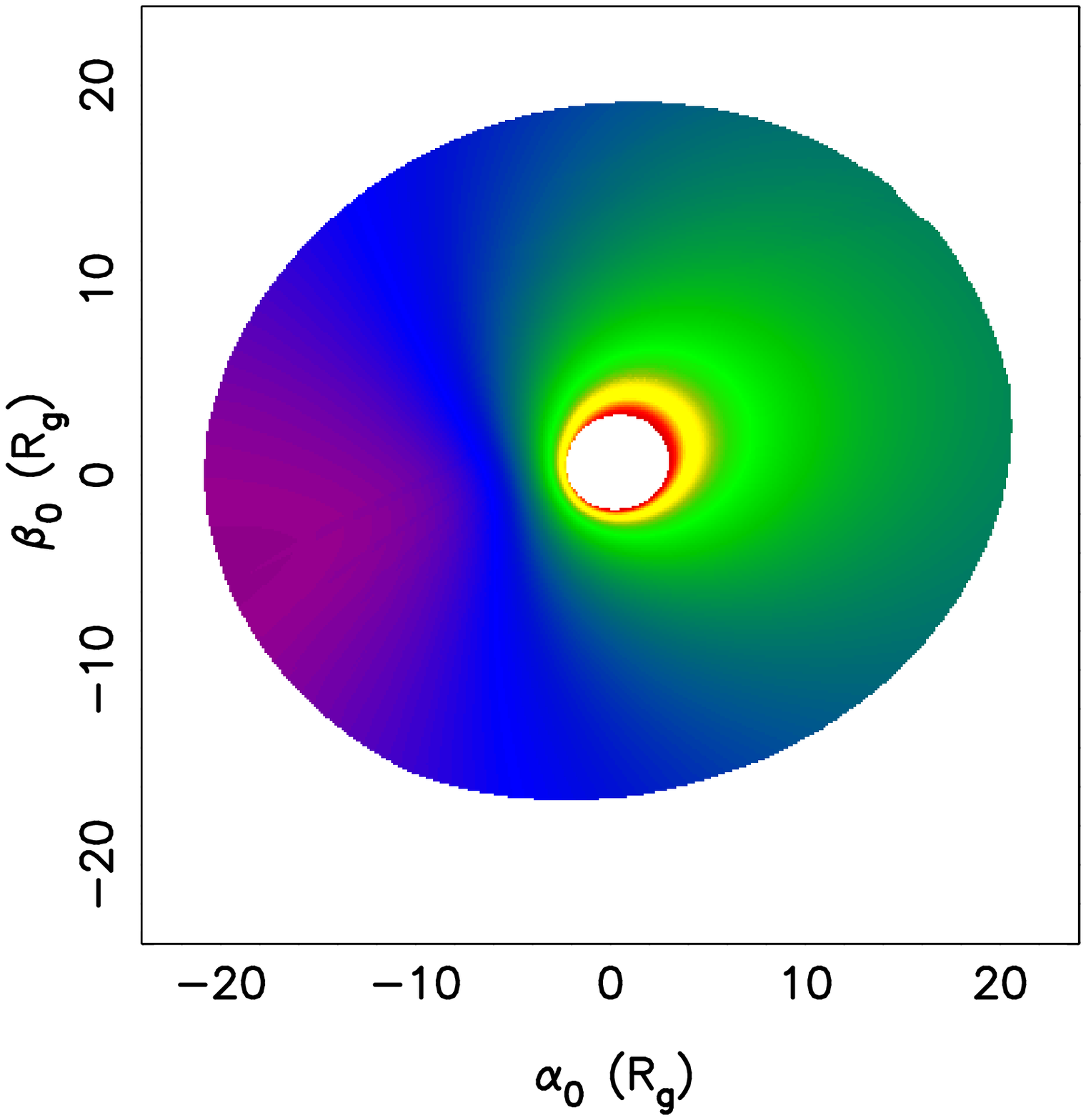}
\includegraphics[height=7cm,width=6.5cm,trim=0.0cm 0.0cm 0.0cm
 0.0cm,clip=true]{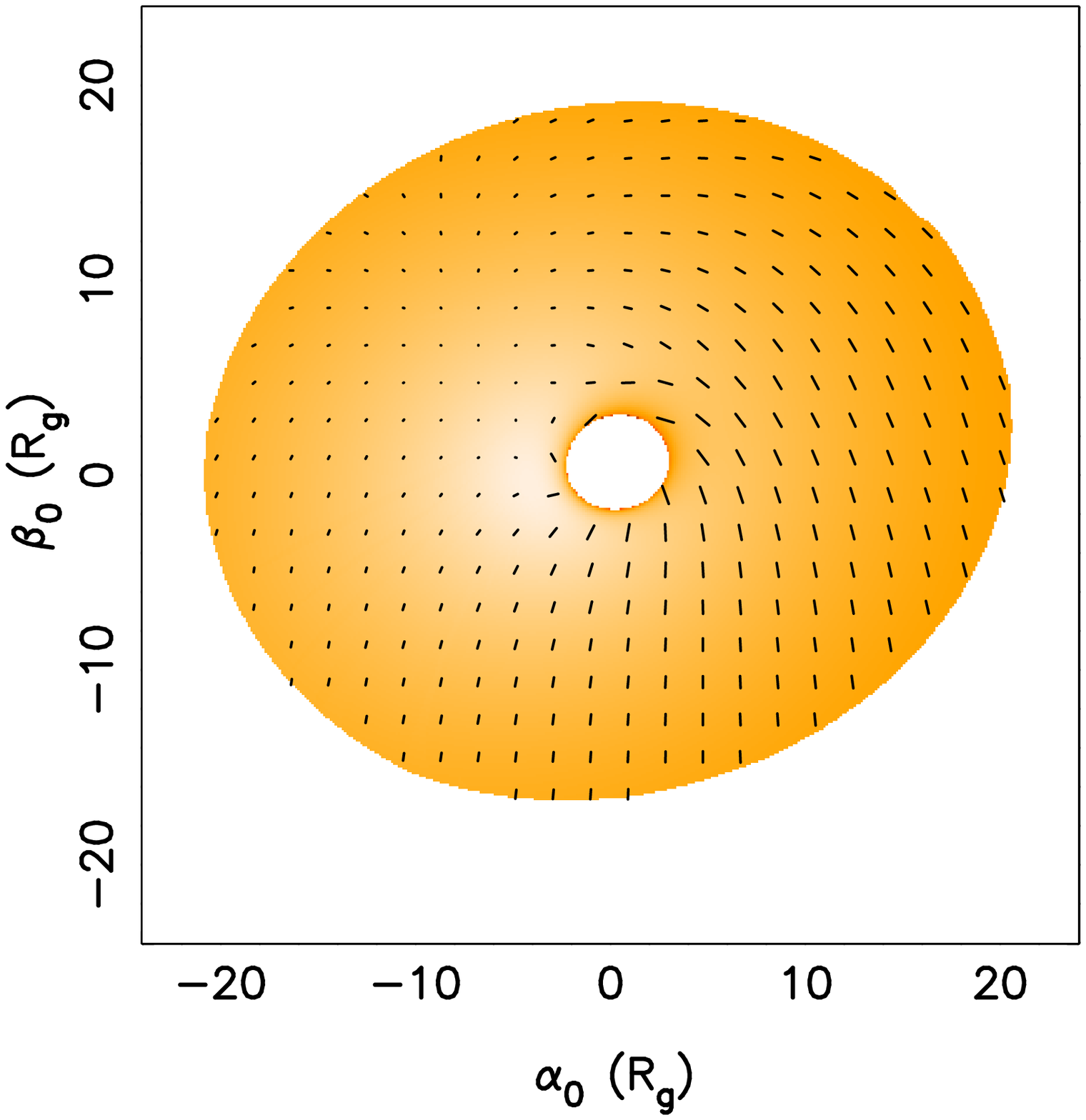} \\
\includegraphics[height=7cm,width=6.5cm,trim=0.0cm 0.0cm 0.0cm
 0.0cm,clip=true]{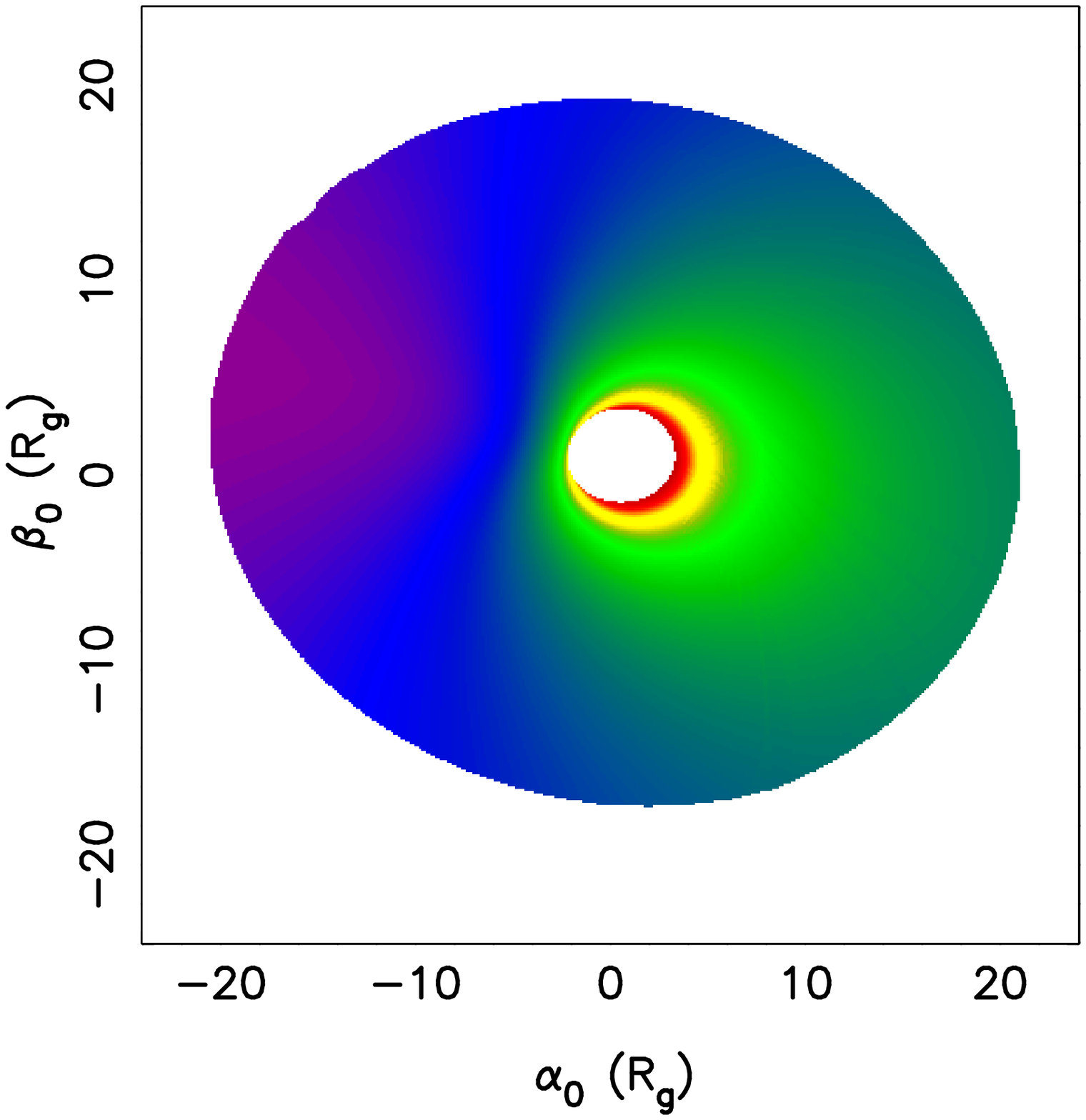}
\includegraphics[height=7cm,width=6.5cm,trim=0.0cm 0.0cm 0.0cm
 0.0cm,clip=true]{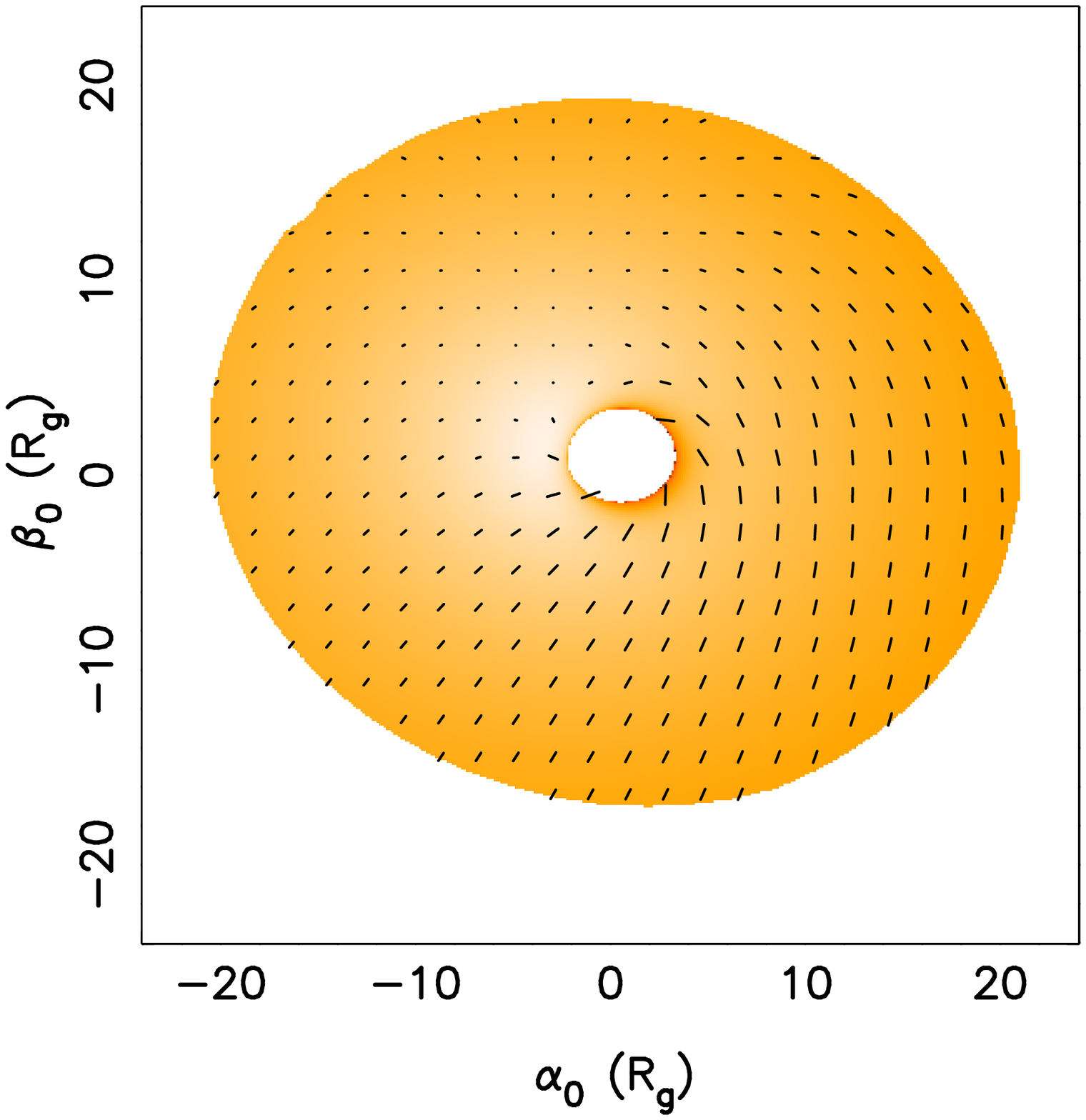} \\
~~~~~~~\includegraphics[height=5.5cm,width=1.5cm,trim=-0.3cm 0.6cm 0.0cm
 0.0cm,clip=true, angle=-90]{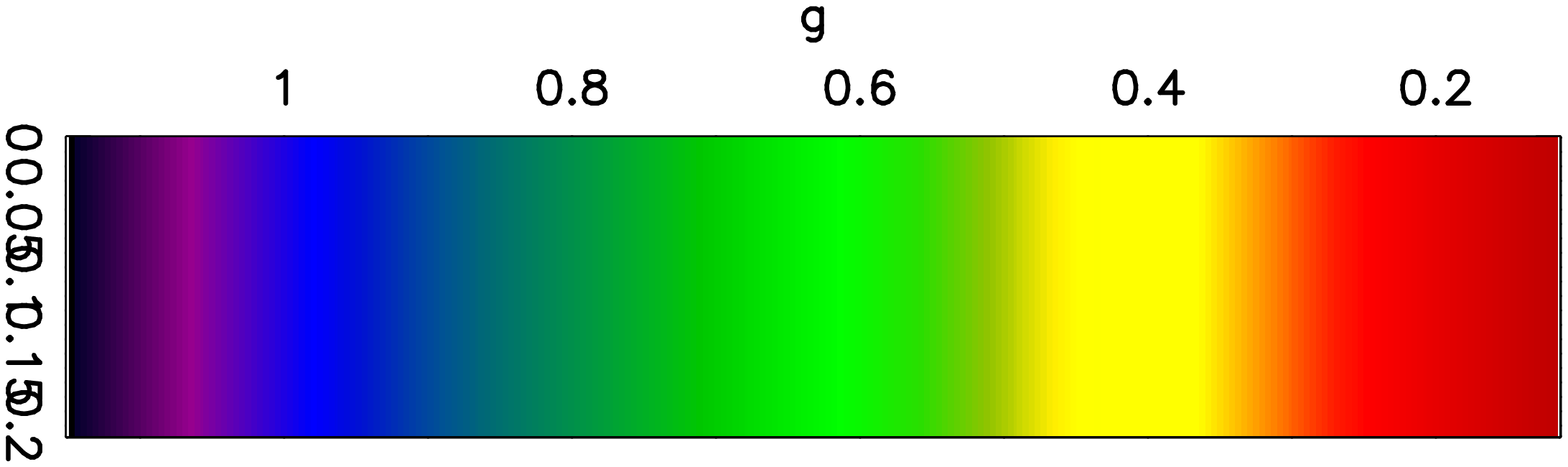} ~~~~~~~~~~~~~~~~
\includegraphics[height=5.5cm,width=1.5cm,trim=0.0cm 0.6cm 0.0cm
 0.0cm,clip=true, angle=-90]{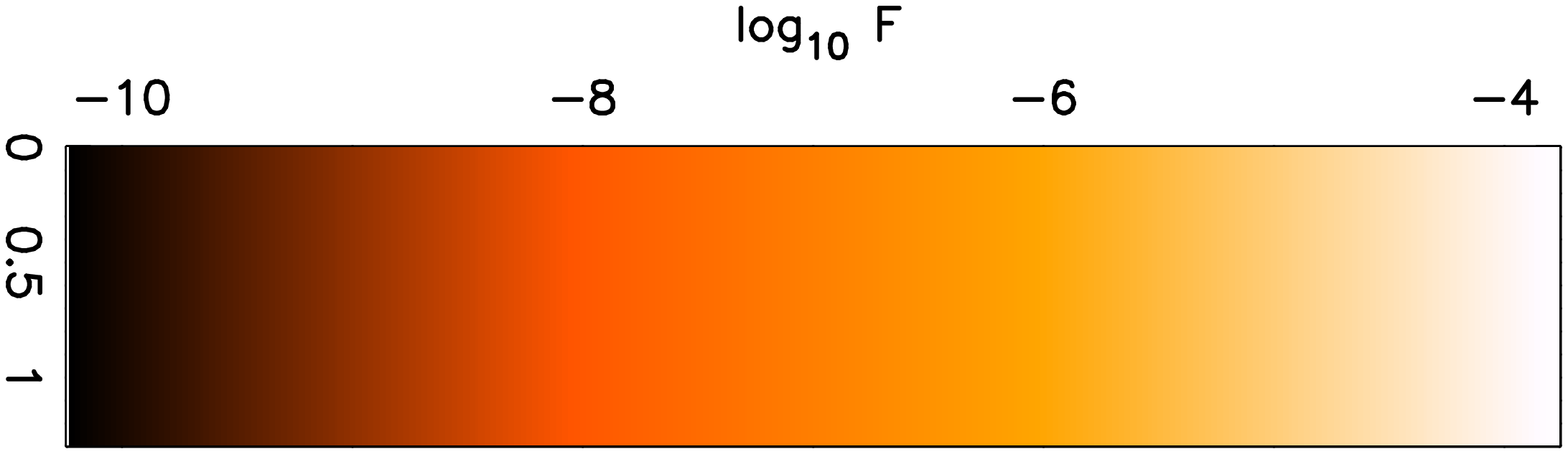}
\caption{Images of the flow for three precession angles for the model
   with $i=30^\circ$, $\Phi=180^\circ$, $\beta=10^\circ$ and
   $h/r=0.1$. The left hand side pictures blueshift and the right hand
   side pictures flux with the polarization vector overlaid,
   normalised to the maximum observed polarization degree. The three
   precession angles pictured, in units of cycles, are $\omega=0$,
   $0.3125$ and $0.625$ from top to bottom respectively. Full movies
   corresponding to these images can be found at {\footnotesize \color{blue}
  \underline{\smash{http://figshare.com/articles/Polarization\_modulation\_gifs/1351920}}
  \color{black}}. Individual movies can alternatively be found on
YouTube: {\footnotesize \color{blue}
  \underline{\smash{www.youtube.com/watch?v=TSe-{}-iXofu8\&feature=youtu.be}}
  \color{black}} (blueshift) and
{\footnotesize \color{blue}
  \underline{\smash{www.youtube.com/watch?v=GjlIRfkor\_s\&feature=youtu.be}}
  \color{black}} (flux and polarization).}
\label{fig:i30an}
\end{figure*}

In our model, the geometry of the flow is governed by three
parameters: the inner radius $r_{\rm i}$, the outer radius $r_o$ and the
scale height $h/r$. In this Paper, we consider parameters appropriate
for the HIMS and so fix $r_o=20$ and $h/r=0.1$
(e.g. \citealt{Ingram2011}; \citealt{Ingram2012}). We set $r_{\rm i}=r_{\rm
  ISCO}$ and assume $a=0.98$. This assumption of high spin maximizes the
impact of relativistic effects which tend to wash out variability,
therefore yielding conservative estimates for the QPO amplitudes
predicted by the model. This corresponds to a QPO frequency of $\sim
2$ Hz, typical of the HIMS. We additionally fix the misalignment angle to
$\beta=10^\circ$ following VPI13 and assume a spectral index of
$\Gamma=2$, again typical of the HIMS.

\subsection{High and low inclination examples}
\label{sec:highandlow}

We consider the observational appearance of the specific geometrical
setup described above for a range of viewing angles. Since our
geometry is asymmetric, the viewer must be specified by both a polar
angle $i$ and an azimuthal angle $\Phi$. In this section, we first
explore two specific examples, chosen to represent respectively
high and low inclination sources. For the high inclination model, we use
$i=70^\circ$ and $\Phi=110^\circ$. For the low inclination model, we
use $i=30^\circ$ and $\Phi=180^\circ$. We then move on to explore a
grid of viewing angles in the following subsection. For these specific
examples, we use a high resolution with 400 steps in the impact
parameters $b$ and $\varphi$ (i.e. $400\times 400$ pixels) and we
consider 32 precession angles.

Figure \ref{fig:i70an} shows images of the flow for three values of
the precession angle. The right hand images depict flux, with colors
defined by the key beneath in arbitrary units. We see the
characteristic apparent warping of the flow through light bending,
with the back of the flow appearing to bend towards us. Since the BH
is spinning rapidly, we also see an asymmetry to this apparent warping
from the frame dragging effect. In the top and bottom plots, we see
the underside of the flow due to rays bending dramatically from their
starting point before passing through the gap between the flow and the
disk and towards the observer. In reality, we may expect material in
some transition region between the two accretion regimes to block our
view of this secondary image to some extent. In these images,
precession and rotation are anti-clockwise. We see that the brightest
patch of the flow is always just to the left of the BH. This region
not only has the largest radial emissivity, but emission from here is
Doppler boosted by the rapid rotation of material in the flow. Also,
parts of the flow are blocked from view by the outer disk. This can be
seen for all three precession angles but is perhaps clearest in the
top plot where the bottom left corner of the flow is hidden behind a
disk which is not pictured.

The images on the left show blueshift for the same three precession
angles, with a key again included beneath. Emission from the
approaching material to the left of the BH is blue shifted by Doppler
effects, whereas the receding material to the right of the BH is red
shifted. Close to the BH, we see the effects of gravitational red
shift, whereby photons lose a significant amount on energy escaping
the gravitational pull of the BH. For selected pixels on the right
hand plot, we also represent the polarization vector with a
black line. The length of the line gives the magnitude of the
polarization degree, normalised to the maximum measured from the
entire run. The orientation of these vectors is heavily influenced by
GR through two main effects (\citealt{Stark1977}). First of all, light
bending means that photons reaching the observer may have had a
different trajectory upon emerging from the flow and therefore the
polarization vectors started off misaligned with one
another. Secondly, the parallel transport of the polarization vector
in heavily curved space-time further changes the orientation. These
effects are, as expected, stronger for photons which passed closer to
the BH, including photons from the back of the flow that emerged from
relatively large radii but needed to pass very close to the BH to
reach the observer. We also see an asymmetry in 
this effect which is directly due to the frame dragging effect. The
orientation of this vector oscillates as a function of precession
angle because the orientation of the flow itself is oscillating. This
is diluted by GR effects but not entirely.
%The full movies for the
%images shown in the left and right panels of Figure \ref{fig:i70an} can be found online at {\small \color{blue}
%  \underline{\smash{www.youtube.com/watch?v=Q2CwOGKVC9U\&feature=youtu.be}}
%  \color{black}} and {\small
%  \color{blue}\underline{\smash{www.youtube.com/watch?v=E3kYAnS3pQI\&feature=youtu.be}} 
%  \color{black}} respectively.

Figure \ref{fig:i70stoke} shows the (normalised) integrated flux,
polarization degree and polarization angle plotted against QPO phase
for this run of the model, with solid and dashed lines representing
respectively the full calculation and the simplified Newtonian
approximation. We see that relativistic effects wash out variability
in the flux and the polarization signature as well as reducing the
average observed polarization degree and angle. The amplitude of
variability is reduced mainly through light bending, which allows us
to see the back of the flow even when the angle between the flow spin
axis and the line of sight is large. The overall polarization degree
is lower for the full calculation because the polarization vectors
observed from different regions of the flow are not aligned, and
therefore do not add together completely constructively as in the
Newtonian approximation. GR influences the polarization angle because
the polarization vectors of emergent photons appear to be bent around
the BH. The vectors in Figure \ref{fig:i70an} to the left of the BH
are forced clockwise and those to the right are forced
anti-clockwise. This is a more pronounced effect on the left hand side
of the BH because the `critical point', where only light rays that are
emitted perpendicular to the flow mid-plane can reach the observer, is
situated here (due to frame dragging). At this point, the observer
sees the polarization vector rotate, mainly due to special
relativistic aberration (see e.g. Figure 3 in \citealt{Dovciak2010}).

For the full calculation, the flux has a fractional rms of
$10.0\%$ and $2.24\%$ in the first and second harmonics
respectively. This is representative of QPOs in high inclination
sources (e.g. GRS 1915+105; \citealt{Yan2013}) for the first harmonic
but the amplitude of the second harmonic is slightly lower than is
typically observed. For the polarization signature, we instead
consider the \textit{absolute} rms, since this is relevant for
detection. This is respectively $1.0\%$ and $0.3\%$ for
the first and second harmonics of the polarization degree and
respectively $3.0^\circ$ and $0.2^\circ$ for the first and second
harmonics of the polarization angle. Also, the polarization degree
lags the flux by $147.3^\circ$ for the first harmonic but the polarization
angle leads the flux by $125.4^\circ$. To estimate the
  importance of the secondary image, we additionally perform a
  calculation whereby rays that pass under the disk plane and
  hit the underside of the flow are assumed to be blocked. For this
  alternative calculation, the flux has a slightly higher fractional
  rms of $13.3\%$ and $3.75\%$ for the first and second harmonics
  respectively. This is because the secondary image appears to have a
  roughly constant shape to the observer and so serves to wash out
  variability. The polarization properties are remarkably similar for
  the alternative calculation: the polarization degree has an rms of
  $1.0\%$ and $0.3\%$ for the first two harmonics and the angle has an
  rms of $3.1^\circ$ and $0.1^\circ$.

Figure \ref{fig:i30an} shows images of the low inclination model, for
the same three precession angles as previously depicted in Figure
\ref{fig:i70an}. For both the flux (right) and blueshift (left), we
still see Doppler shifts due to rapid rotation, gravitational redshift
and the effects of light bending, however all these effects are less
pronounced since the component of velocity in the line of sight is
lower and also less photons need to pass close to the BH in order to
reach the observer.
%The full movies for the
%images shown in the left and right panels of Figure \ref{fig:i30an}
%can be found online at
%{\small \color{blue}
%  \underline{\smash{www.youtube.com/watch?v=TSe-{}-iXofu8\&feature=youtu.be}}
%  \color{black}} and
%{\small \color{blue}
%  \underline{\smash{www.youtube.com/watch?v=GjlIRfkor\_s\&feature=youtu.be}}
%  \color{black}} respectively.
The full movies for all of the images shown in Figures \ref{fig:i70an}
and \ref{fig:i30an} can be found at {\small \color{blue}
  \underline{\smash{http://figshare.com/articles/Polarization\_modulation\_gifs/1351920}}
  \color{black}} (additionally, YouTube links for individual movies
are given in the Figure captions).

Figure \ref{fig:i30stoke} shows the flux, polarization
degree and polarization angle plotted against QPO phase for this
model, again with the results of the simple Newtonian calculation
represented with dashed lines. Here, the flux is less affected by
light bending, since the angle between the flow spin axis and the line
of sight never gets particularly large. GR effects only reduce the
mean and rms of the polarization degree and angle, with the phase of
the oscillations only slightly modified from the Newtonian
approximation. The amplitude of flux variability is lower here than in
the high inclination case, although the inclination dependence of
amplitude is far more pronounced for the Newtonian approximation than
for the full calculation. This is again down to light bending. The
absolute rms of the polarization degree is comparable to the high
inclination case, but the \textit{fractional} variation in polarization
degree is actually significantly larger here than for the high
inclination model. This can be understood from Figure \ref{fig:polmu}
which shows that the emergent  polarization degree is a far steeper
function of $\mu_e$ for large $\mu_e$. However, the mean polarization
degree is far lower here than for the high inclination angle, and
these two considerations happen to balance. The amplitude of the
polarization angle oscillation is greater for the low inclination
model, which is purely a geometrical effect. It is clear from Figure
\ref{fig:i30stoke} that the polarization degree and flux are nearly in
phase for this example, in contrast to the results for the high
inclination model. We find that $p$ lags $F$ by $20.3^\circ$ for the
fundamental and $\chi$ lags $F$ by $113.8^\circ$. Since we see
  no secondary image from the underside of the flow for this viewing
  angle, assuming the material between the disk and flow to be
  optically thick does not change our results.

\subsection{Parameter study}
\label{sec:param}

\begin{figure*}
\centering
\includegraphics[height=10cm,width=8cm,trim=0.0cm 0.0cm 0.0cm
 0.0cm,clip=true]{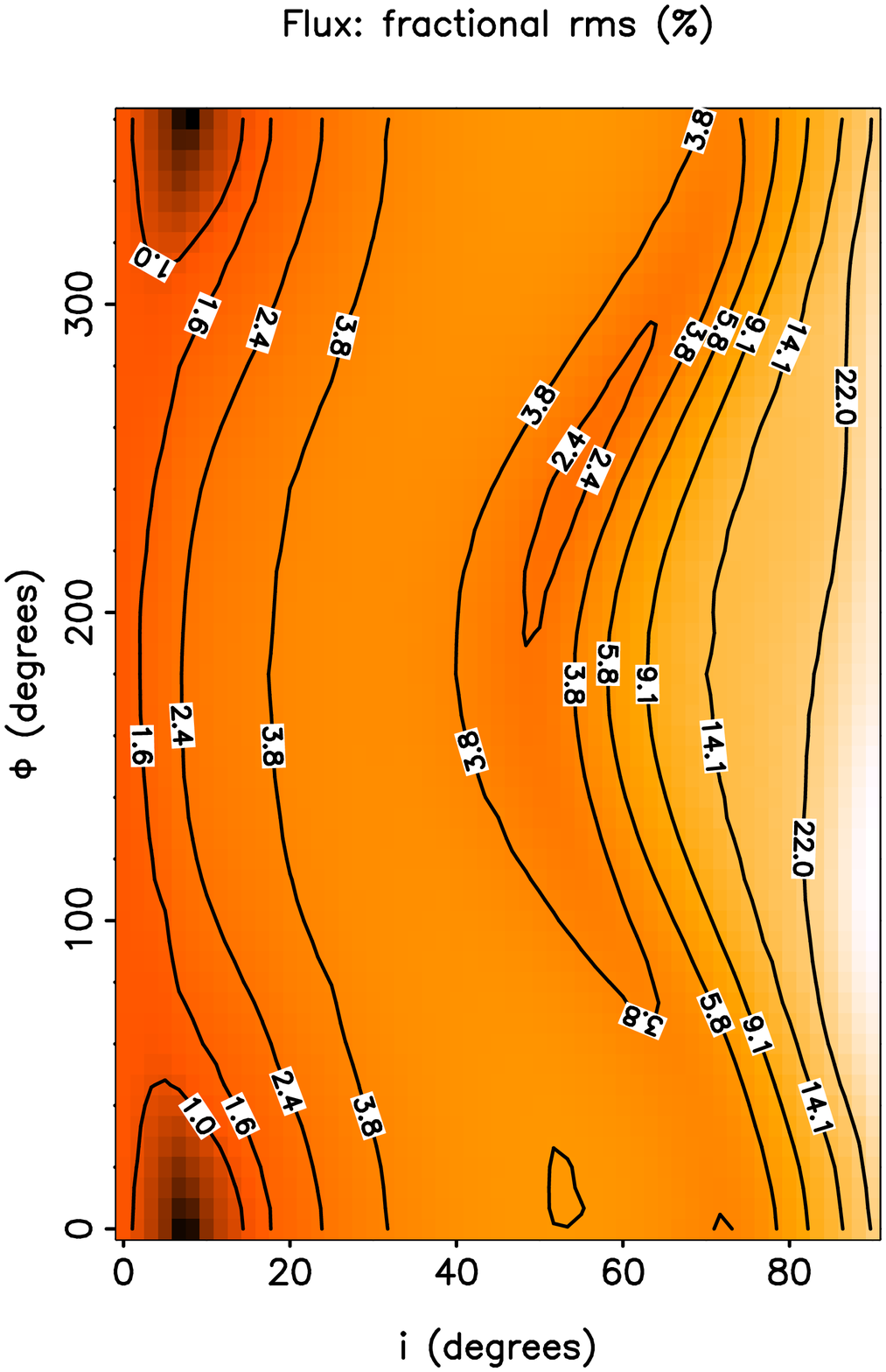}
\includegraphics[height=10cm,width=8cm,trim=0.0cm 0.0cm 0.0cm
 0.0cm,clip=true]{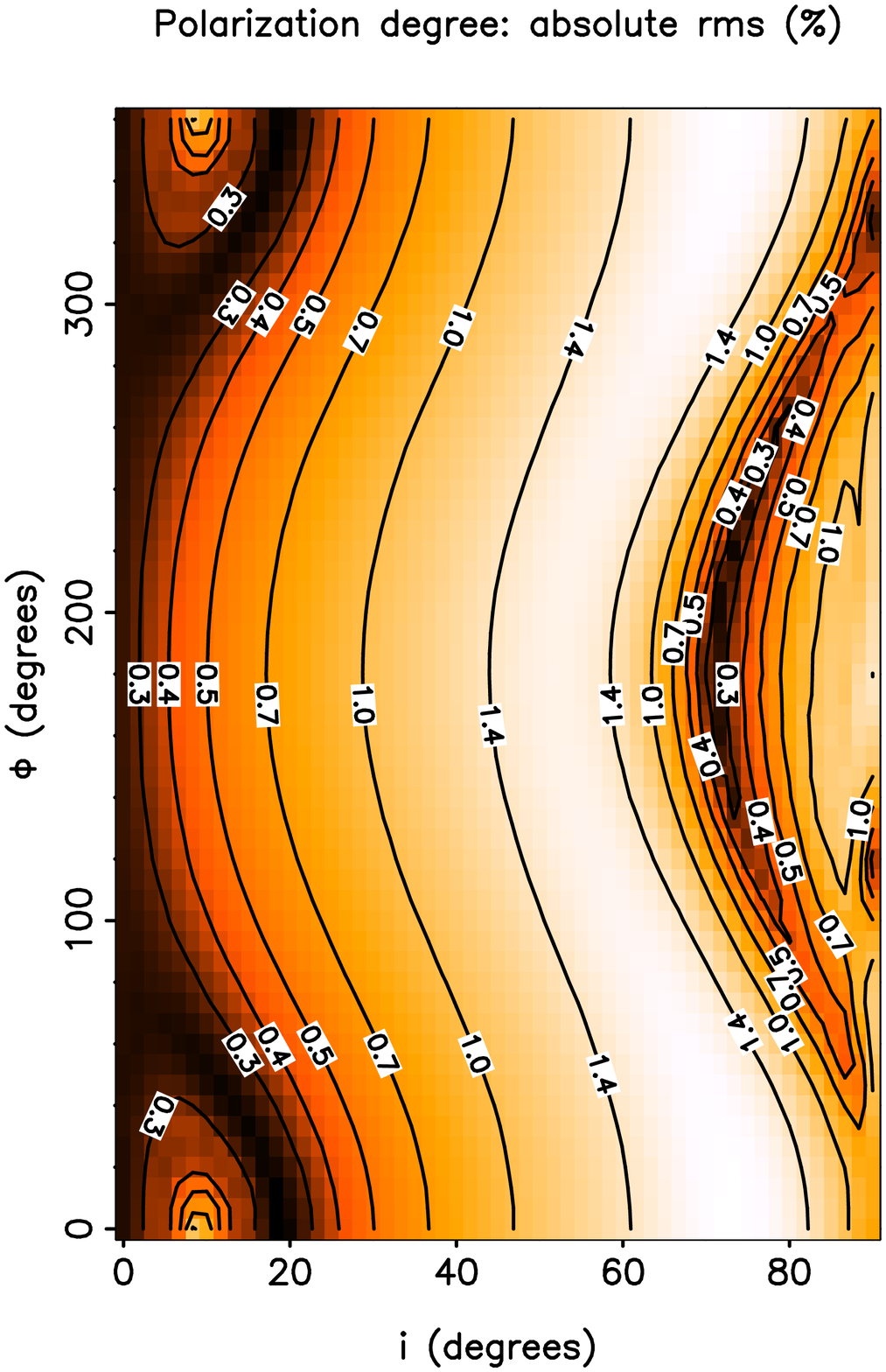} \\
\vspace{5mm}
\includegraphics[height=10cm,width=8cm,trim=0.0cm 0.0cm 0.0cm
 0.0cm,clip=true]{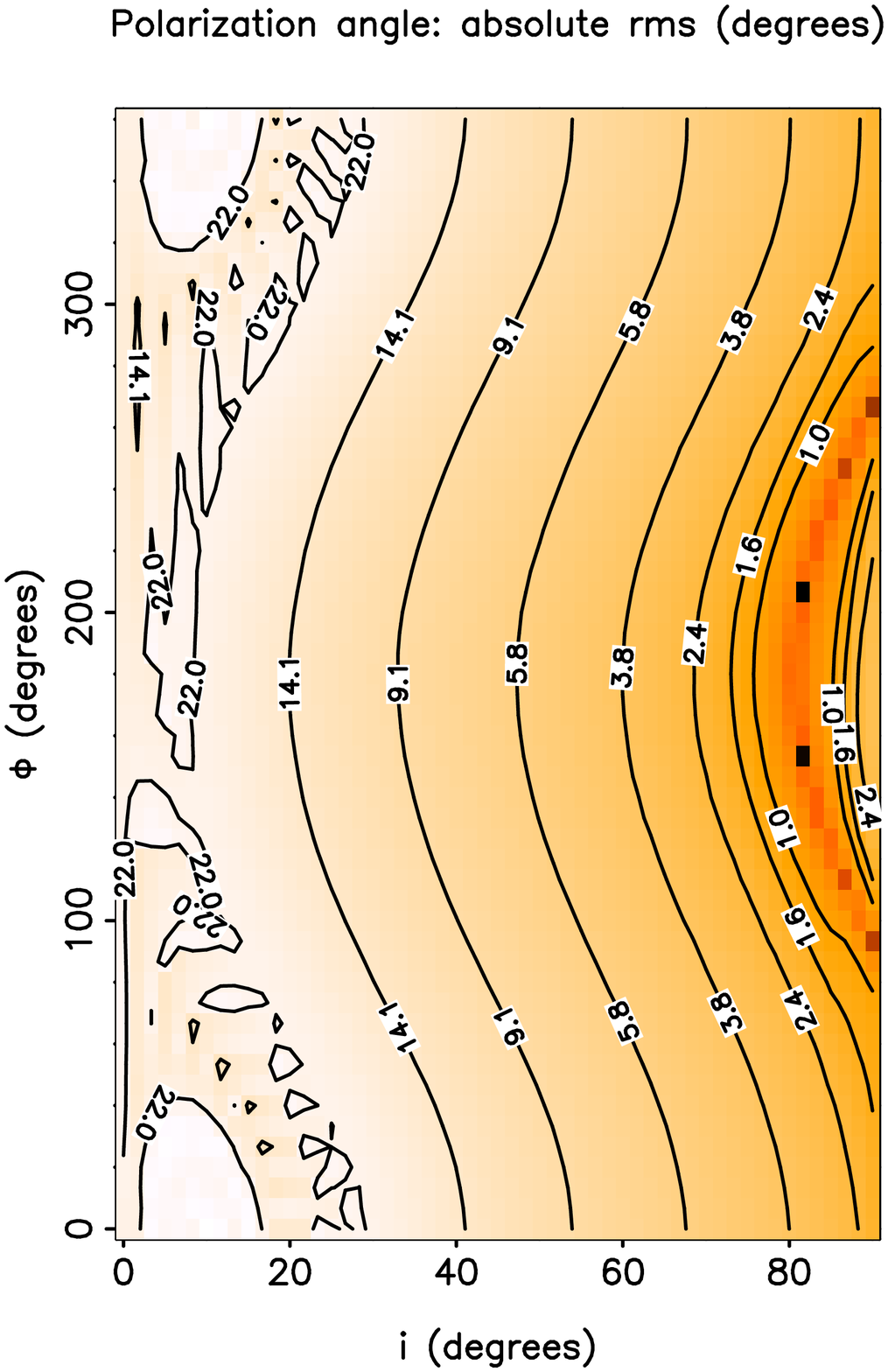}
\includegraphics[height=10cm,width=8cm,trim=0.0cm 0.0cm 0.0cm
 0.0cm,clip=true]{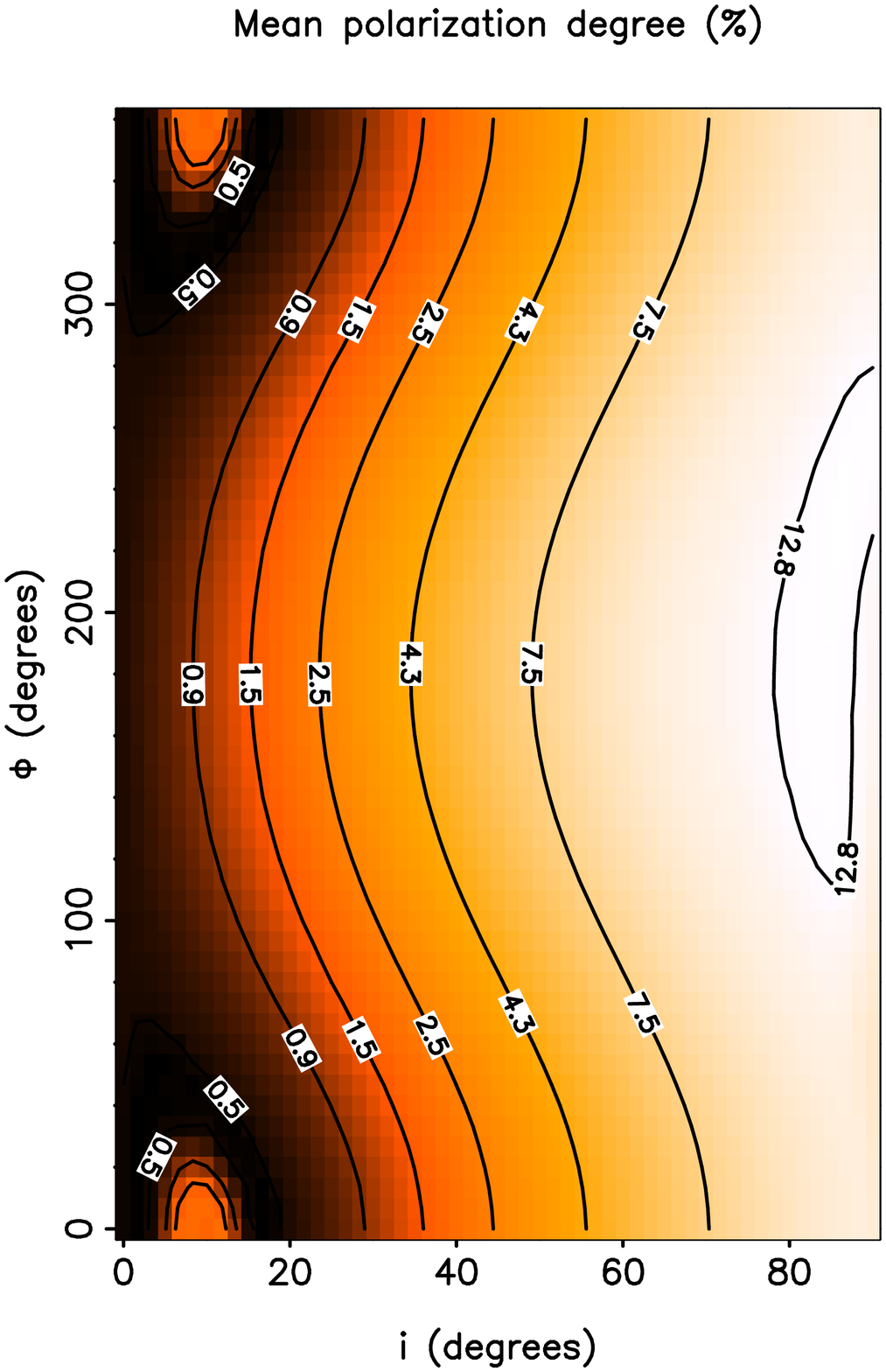}
\caption{\textit{Top-Left:} Fractional rms in the first harmonic of
  the flux modulation plotted as a function of the two viewing angles,
  $i$ and $\Phi$, with contours labelled as
  percentages. \textit{Top-Right:} Absolute rms in the first harmonic 
  of the polarization degree modulation, with contours again labelled
  as percentages. \textit{Bottom-Right:} Mean polarization degree,
  with contours once again labelled as
  percentages. \textit{Bottom-Left:} Absolute rms in the first
  harmonic of the polarization angle modulation, with contours now
  labelled in degrees. The colors in each plot follow separate
  logarithmic scales. We see that the amplitude of the flux and
  polarization degree modulations increase with inclination angle,
  $i$, whereas the amplitude of the polarization angle modulation
  reduces with $i$. Note the slight asymmetry in all plots around
  $\Phi=180^\circ$, which occurs because $a \neq 0$.}
\label{fig:rms}
\end{figure*}

We now consider the full range of viewing angles for the specific
geometrical setup described at the start of this section. The
corresponding parameter exploration in VPI13 considered viewing
angles ranging from $0^\circ \leq i \leq 90^\circ$ and $0^\circ \leq
\Phi \leq 180^\circ$. However, our use of the Kerr metric introduces a
subtle asymmetry (which disappears for $a=0$ of course) meaning that
we must explore the range $0^\circ \leq \Phi \leq 360^\circ$ to be
exhaustive. We use a reduced resolution of $80\times 80$ pixels and
consider $16$ precession angles. When we test for the specific high
and low inclination models considered above, we find that this reduced
resolution provides a very good approximation to the high resolution
run.

\begin{figure*}
\centering
\includegraphics[height=10cm,width=8cm,trim=0.0cm 0.0cm 0.0cm
 0.0cm,clip=true]{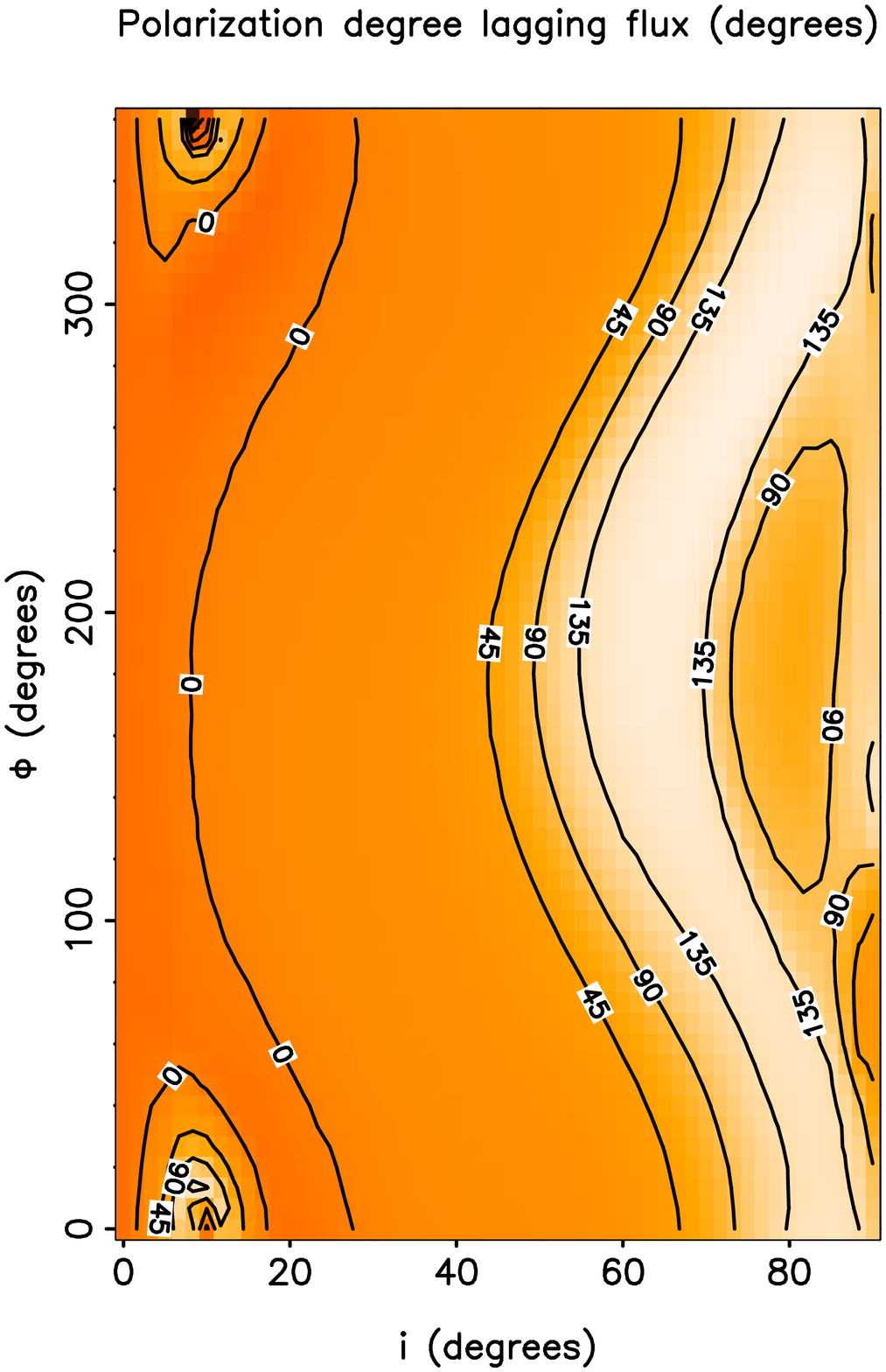} ~~~~
\includegraphics[height=10cm,width=8cm,trim=0.0cm 0.0cm 0.0cm
 0.0cm,clip=true]{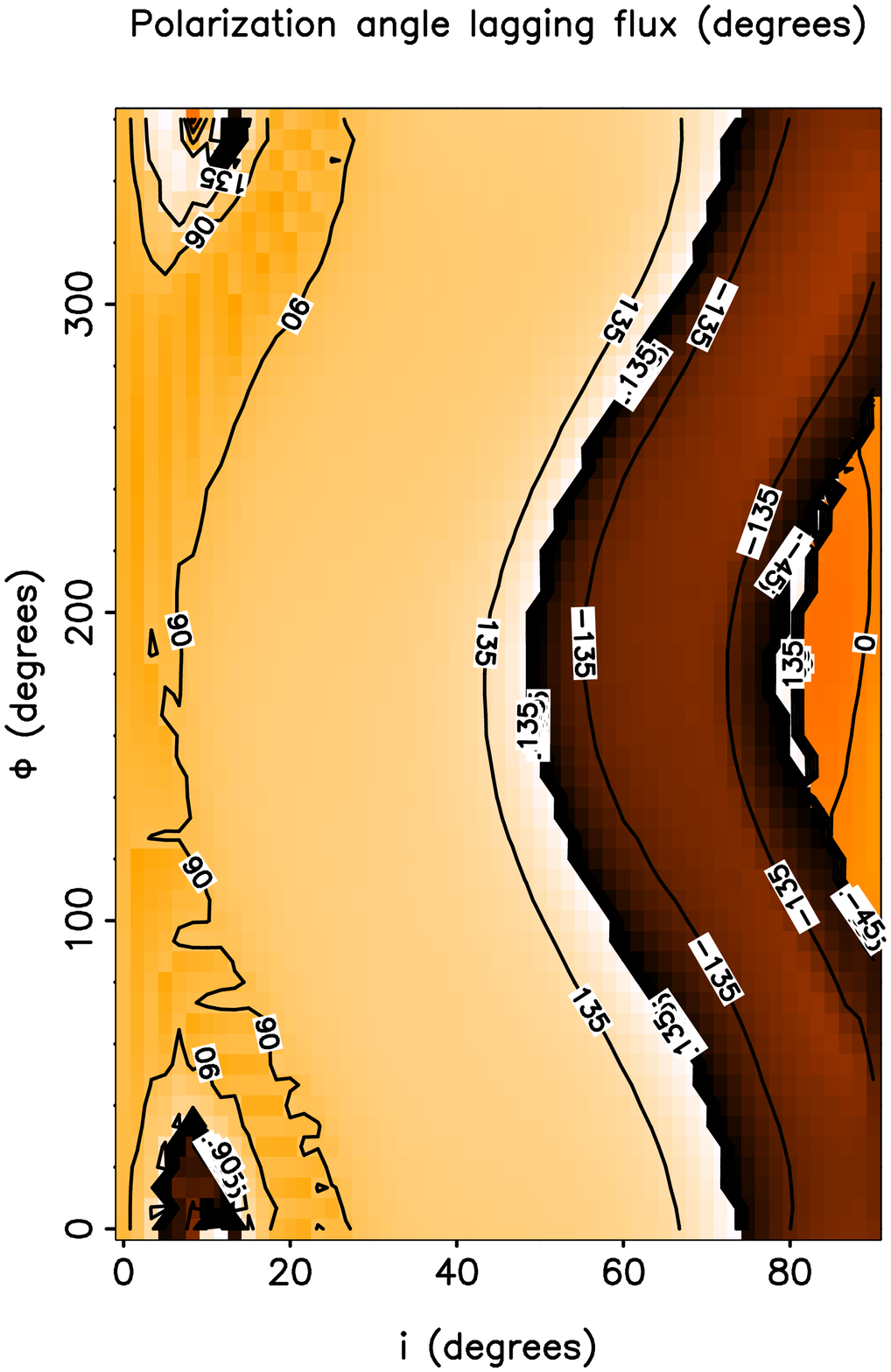}
\caption{\textit{Left:} Phase lag in degrees between the polarization
  degree and the flux, with positive lag meaning that $p$ lags
  $F$. Here, we only plot for the fundamental and color scales
  linearly from $-180^\circ$ (darkest) to $180^\circ$ (lightest). We
  see that the model predicts the oscillation in polarization degree
  to lag the first harmonic of flux for almost the full range of
  viewing angles. \textit{Right:} Phase lag between polarization
  angle and flux. The model predicts the magnitude of the phase lag
  between $\chi$ and $F$ to be large for most of parameter space.}
\label{fig:lag}
\end{figure*}

\begin{figure*}
\centering
\includegraphics[height=10cm,width=8cm,trim=0.0cm 0.0cm 0.0cm
 0.0cm,clip=true]{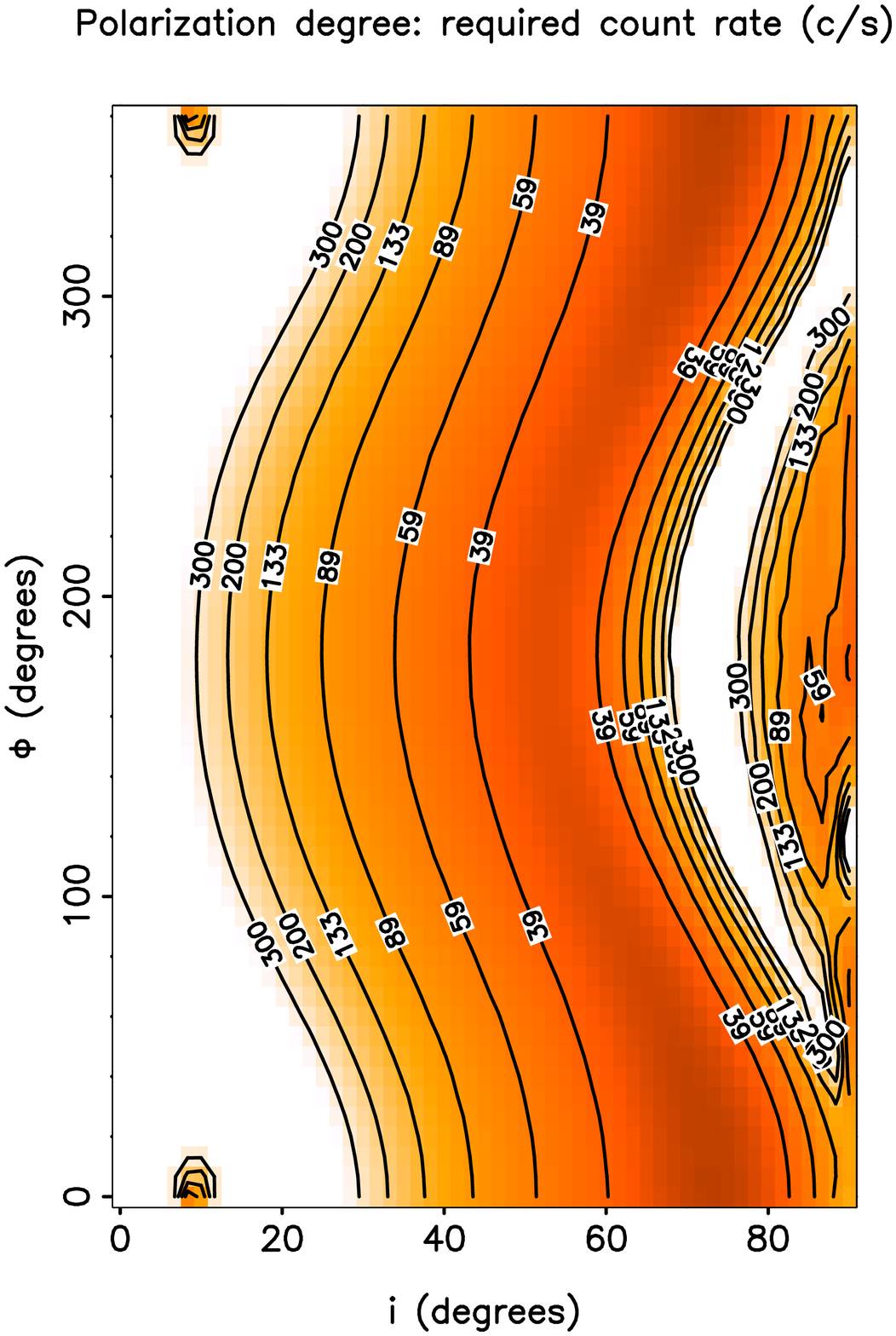} ~~~~
\includegraphics[height=10cm,width=8cm,trim=0.0cm 0.0cm 0.0cm
 0.0cm,clip=true]{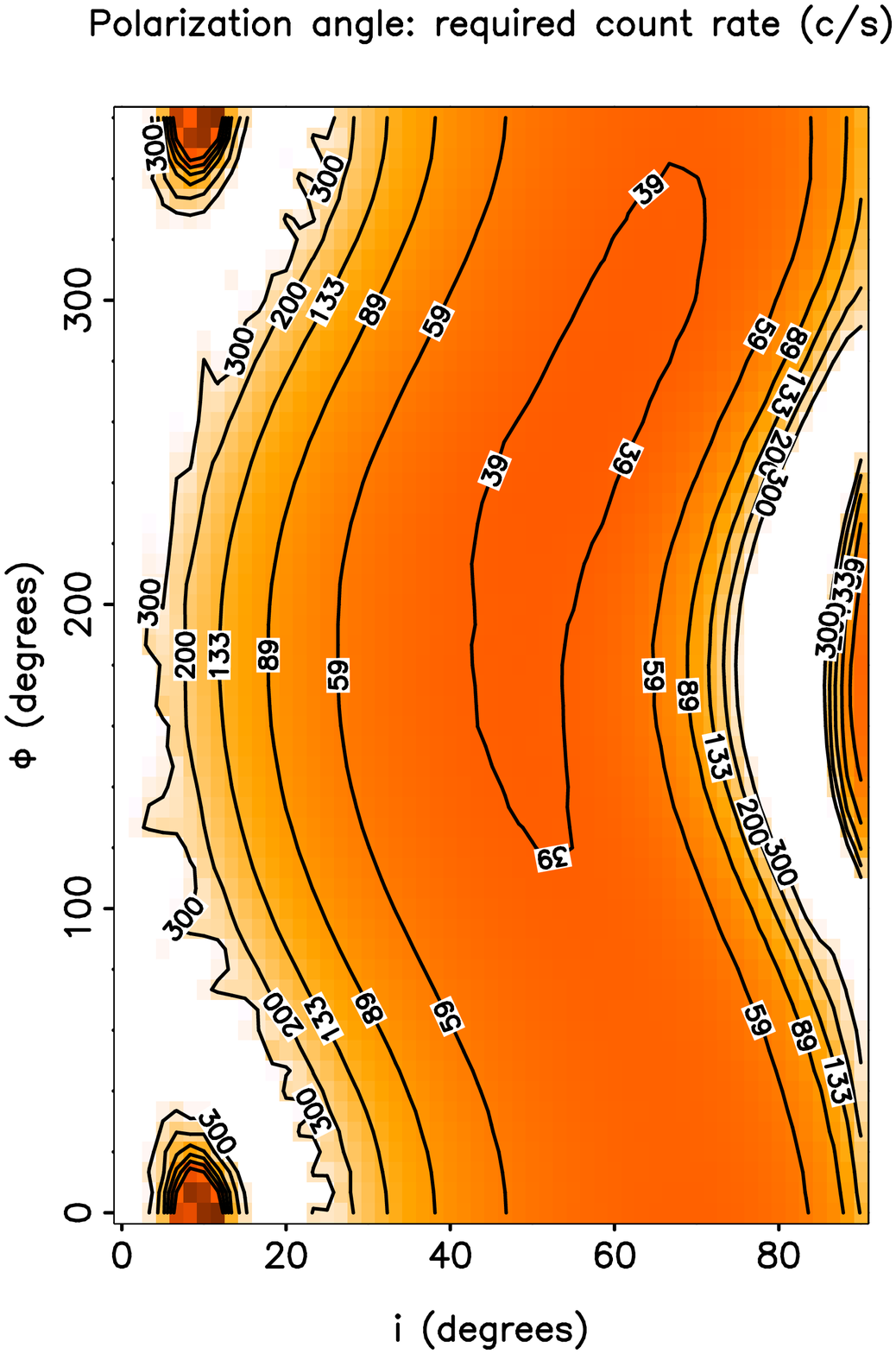}
\caption{Required count rate in order to detect a modulation in
  polarization degree (left) and angle (right). We assume a $200$ ks
  exposure and a modulation factor of $\mu=0.5$. See text for more
  details.}
\label{fig:cps}
\end{figure*}

In Figure \ref{fig:rms}, we use colors and contours to plot four
different quantities for the entire range of viewing angles. Each plot
has it's own logarithmic scale. The top-left plot shows the fractional
rms amplitude of the flux modulation, $\sigma_F/\langle F
\rangle$, for the first harmonic only. The contours in this plot are
labeled as percentages and labels always face in the uphill
direction. We see that the amplitude of the QPO increases with $i$ as
we may expect from the previous subsection. This is consistent with
observations (\citealt{Schnittman2006}; \citealt{Motta2014b};
\citealt{Heil2015}). We note that this rise of amplitude with
  inclination angle would be even more pronounced if we had assumed
  secondary images to be blocked by optically thick material, since
  these wash out variability for high viewer inclinations. The plot
(and all subsequent plots) is nearly symmetric about $\Phi=180^\circ$
but a subtle asymmetry is introduced by the assumption of high
spin. The top-right plot shows the absolute rms amplitude of the
polarization degree modulation $\sigma_p$, again for the fundamental
only, with contours labeled as percentages. This peaks at $i \sim
60-70^\circ$. The bottom-right plot shows mean polarization degree
$\langle p \rangle$, again with contours labeled as percentages. We
see that this increases with $i$. For small $i$, the mean polarization
degree is very low indeed, which leads to the \textit{fractional} rms
amplitude of the polarization degree modulation becoming very large
for low inclinations. However, it is likely the \textit{absolute} rms
amplitude that is relevant to detection.

The bottom-left plot shows the absolute rms variability of
polarization angle $\sigma_\chi$, with contours now labelled in degrees, again for
the first harmonic only. Note that the polarization angle is only
defined on an interval of $180^\circ$, since upward polarization is
indistinguishable from downward polarization. This can produce phase
wrapping: e.g. if $\chi$ oscillates between a minimum and maximum of
$80^\circ$ and $100^\circ$ but is defined on the interval $-90^\circ$
to $90^\circ$, the measured $\chi$ will jump from $90^\circ$ to
$-90^\circ$ at some point during the QPO cycle, producing a spuriously
large rms measurement. To avoid this, for every combination of $i$ and
$\Phi$, we define $\chi$ for the first QPO phase on the interval $-90^\circ$
to $90^\circ$. For each subsequent value of QPO phase, we first
calculate $\chi$ on the same interval, but also try adding $0$,
$180^\circ$ and $-180^\circ$, and choose the interval that minimises
the difference to the previous measurement of $\chi$. We see that the
amplitude decreases with inclination angle, but is above $1^\circ$ for
most of parameter space.

In Figure \ref{fig:lag} (left), we plot the phase lag between polarization
degree and flux against $i$ and $\Phi$. Here, positive lags mean that
$p$ lags $F$ and we again consider only the fundamental. We see that
$p$ lags $F$ for nearly all viewing angles. In the right panel, we
instead plot the phase lag between $\chi$ and $F$. Here we see phase
wrapping at $i \sim 60^\circ$ which occurs because the lag between the
two functions is only defined on the interval $-180^\circ$ to
$180^\circ$. The magnitude of the lag is large for most of parameter
space. This can be understood by looking at the images in Figures
(\ref{fig:i70an}) and (\ref{fig:i30an}). The peak in $\chi$ for our
coordinate system occurs approximately when we see the flow spin axis
tilted the furthest to the left. The flux rarely peaks close to this
phase in the precession cycle. If a high enough signal to noise can be
achieved to observe these lags, they may provide a powerful extra
diagnostic.

\section{Discussion}
\label{sec:discussion}

We have calculated the polarization signature predicted by the
precessing inner flow model for low frequency QPOs. We find that the
polarization degree and angle are expected to be modulated on the QPO
period.

\subsection{Assumptions}

In our analysis, we calculate the GR effects very accurately but make
a number of simplifying assumptions about the properties of the inner
accretion flow. We use analytical parameterizations for the angular
dependence of intensity and polarization degree of radiation emerging
from the flow. Since these parameterizations are based on the
calculations of ST85, this is a reasonable assumption but there
is scope to extend this work in future. Our parameterization is only
valid for one value of optical depth ($\tau=1$) and so we cannot
explore the dependence of our results on this parameter. In addition,
we assume that the angular dependence of the flow emissivity can be
separated from the radial and energy dependencies (Equation
\ref{eqn:sep}). In reality, it is likely that the shape of the emitted
spectrum depends on both radius \textit{and} viewing angle. Since the
outer part of the flow is illuminated by a greater flux of cool disk
photons, the spectrum should be softer for larger $r$, giving rise to
the observed time lags (\citealt{Kotov2001}; \citealt{Ingram2013}).
There is now observational evidence that the spectral shape of the
flow depends on viewing angle (\citealt{Heil2015};
\citealt{Ingram2015}). Similarly, the angular dependence of
polarization degree will likely depend on energy of the emitted
photon, and also perhaps the radius it was emitted from. We also
expect the flow optical depth, and therefore mean polarization degree,
to depend on truncation radius / spectral state. For a low luminosity
hard state we expect $\tau$ to drop below unity and polarization
degree to reach $\sim 50\%$ (see \citealt{Viironen2004}). Replacing
our simple parameterization with a full Monte Carlo simulation in
future will be a significant improvement upon this work.

We also effectively assume that the seed photon luminosity stays
constant during a QPO cycle. This is a good assumption if the seed
photon luminosity is completely dominated by internally generated
photons (\citealt{Veledina2011}; VPI13). However, the luminosity of
disk photons incident on the flow will vary as the misalignment
between the disk and flow changes over the course of a precession
cycle. A calculation of disk seed photon luminosity as a function of
precession angle has yet to be performed taking GR effects fully into
account. The relative importance of internally generated seed photons
will increase with truncation radius. It should be possible to put
constraints on such an evolution of the seed photon origin with X-ray
polarimetry, since polarization degree depends strongly on scattering
order for disk seed photons (see Fig. 2 in \citealt{Viironen2004}).

\subsection{Detection}

It may be possible to detect the predicted polarization degree and
angle modulation in the near future with a dedicated polarization
satellite mission. Here we present a simple calculation to roughly
assess detectability. X-ray polarimeters generally use either Thomson
scattering (e.g. the \textit{Polarization Spectroscopic Telescope
  Array}; \textit{PolSTAR}) or the photoelectric effect (e.g.
\textit{Gravity \& Extreme Magnetism SMEX}; \textit{GEMS}). Scattering
polarimeters measure the landing position of scattered photons, which
travel preferentially in the direction perpendicular to their electric
field vector (e.g. \citealt{Guo2013}). Photoelectric effect
polarimeters track the direction of photoelectrons, which are
preferentially emitted in the direction parallel to the incoming electric
field vector (e.g. \citealt{Black2010}). In either case, an estimate
for the polarization angle of each incident photon, $\chi_k$ (i.e. for
the $k^{\rm th}$ photon), can be recorded. When many photons are
incident on the detector, a polarized signal will exhibit a sinusoidal
modulation with distribution (\citealt{Henric2011}; \citealt{Lei1997};
\citealt{Kislat2014}): 
\begin{equation}
f( \chi_k) \propto 1 + p~\mu \cos[ 2( \chi_k - \chi ) ],
\end{equation}
where $\chi$ is the `true' polarization angle of the source, $p$ is
the `true' polarization degree and $\mu$ is the modulation factor. The
performance of a polarimeter can be characterised by the modulation
factor, since it governs the distribution of a $100\%$ polarized
signal. When $C$ photons are incident on the detector, the measurement
error on $p$ is (\citealt{Kislat2014})
\begin{equation}
dp \approx \sqrt{ \frac{ 2/\mu^2 - p^2 }{ C-1 } },
\end{equation}
when $C$ is sufficiently large to be in a Gaussian regime (note, $p$
is represented here as a fraction rather than a percentage). The
measurement error on $\chi$ (in radians) is
\begin{equation}
d\chi \approx \frac{1}{p\mu\sqrt{2(C-1)}}.
\end{equation}

Detection of a $\sim 1$ s QPO requires a time resolution $\leq 0.1$
s. Very few photons can likely be collected in such a short time and
therefore a high time resolution time series of individual Stokes
parameters would be very noisy indeed. It is, however, possible to
stack into QPO phase bins (\citealt{Tomsick2001}), resulting in one
folded time series. If we stack into $N$ phase bins, the number of
counts detected per phase bin is simply $C = R~T/N$, where $R$ is the
mean count rate and $T$ is the total exposure time. We can estimate
that detection of a QPO in the polarization degree requires a
measurement error in each phase bin around one fifth of the absolute
rms variability amplitude, $dp \sim \sigma_p /5$. Thus, the mean count
rate we must be able to detect for the observation is
\begin{equation}
R \sim  \frac{25N}{T\sigma_p^2} \left[ \frac{2}{\mu^2} - \langle p \rangle^2 \right].
\label{eqn:Rp}
\end{equation}
We can estimate the count rate required to detect a modulation in
polarization angle in a similar manner. In this case
\begin{equation}
R \sim 4.1\times10^4 \frac{N}{T \langle p \rangle^2 \mu^2 \sigma_\chi^2},
\label{eqn:Rt}
\end{equation}
where $\sigma_\chi$ is in degrees. In Figure \ref{fig:cps}, we show
these required count rates for all of parameter space assuming $N=8$
phase bins, an exposure of $T=200$ ks and a modulation factor of
$\mu=0.5$. The plots on the left and right are for polarization degree
and angle respectively. We see that to detect either modulations at
all, a polarimeter must have the sensitivity to measure a count rate
above $40$ c/s, and a count rate of $\sim 60$ c/s opens up a 
reasonably large fraction of parameter space ($i \sim
40^\circ-75^\circ$) in which both modulations can be detected. The same
folding method can be used to measure lags between the polarization
properties and the flux, although this may require a more challenging
sensitivity.

A number of missions proposed for \textit{NASA's} Small Explorer
Program (SMEX) should be able to detect count rates $\gtrsim 40$
c/s. \textit{PolSTAR} (the satellite incarnation of the balloon
experiment \textit{X-Calibur}; \citealt{Guo2013}) is sensitive to the
$\sim 10-20$ keV energy range in which the Comptonized spectrum
dominates, and would therefore be ideal for this application. The soft
X-ray missions \textit{GEMS} (\citealt{Black2010}) and \textit{the
  Imaging X-ray Polarimetry Explorer} (\textit{IXPE};
\citealt{IXPE2008}) should also be able to measure a similar count
rate, however contribution to the flux from the constant disk
component at softer energies will dilute the variability amplitude of
the polarization signature, making detection more challenging (${\rm
  rms} \approx {\rm rms}_{\rm flow} [ 1 - x_{\rm disk}]$, where
$x_{\rm disk}$ is the fraction of the flux contributed by the
disk). Disk dilution is not a problem in the hard state, but here the
flux is lower and the QPOs are generally not as coherent as in the HIMS.

\subsection{Implications}

Detection of a polarization modulation on the QPO frequency would have
strong implications. First of all, this would confirm that the QPO is
indeed a geometric effect, as is strongly hinted in the literature at
the moment (\citealt{Motta2014b}; \citealt{Heil2015};
\citealt{Ingram2015}). It will also provide a strong test for the
Lense-Thirring QPO model. This is fairly profound in itself, since
Lense-Thirring precession has never been unambiguously observed in the
strong field regime, but this also has implications with regard to
making spin measurements. Current spin measurements from disk spectral
fitting (e.g. \citealt{Kolehmainen2010}; \citealt{Steiner2011}) assume
that the BH and binary spin axes are aligned, in contradiction of the
Lense-Thirring QPO model. Careful modelling of QPO properties can 
be used to estimate the misalignment angle $\beta$ to improve the
spectroscopic measurements and also the very measurement of the QPO
frequency itself gives an orthogonal spin estimate if high frequency
QPOs are also present (\citealt{Motta2014}; \citealt{Ingram2014}).

\section{Conclusions}
\label{sec:conclusions}

We find that the polarization signature emitted from a truncated disk
/ precessing inner flow geometry oscillates on the QPO frequency. The
modulation in polarization degree has an absolute rms which increases
gradually with viewer inclination angle, peaking at $1.5\%$ for $i
\sim 60^\circ$. In contrast, the absolute rms of the polarization
angle modulation is higher for lower inclination angles. For
polarization degree, this inclination dependence is mainly due to the
assumed angular dependence of polarization degree and emissivity for
Compton scattering, which we parameterize to agree with the
calculations of ST85. Although our parameterization is only
appropriate for an optical depth of $\tau=1$, we note that the results
of ST85 are qualitatively similar for optical depths up to $\tau \sim
2.5$. The inclination dependence of the polarization angle depends
mainly on the precessing geometry and GR effects, and so is robust to
our assumptions about the Compton scattering process. Our calculations
here only consider one specific flow geometry, and we explore the full
range of viewing angles. In future, we will also explore the effects
of changing parameters such as the truncation radius and spin. In
particular, we note that assuming a larger misalignment angle,
$\beta$, will increase the predicted amplitudes of all
modulations.

We find through a rough calculation that, in order to detect this
effect for a reasonable fraction of parameter space, an X-ray
polarimeter will need to detect a $10-20$ keV count rate of $\gtrsim
60$ c/s from a bright object displaying QPOs. The current generation
of proposed X-ray polarimetery missions will likely fill this
requirement. In particular, \textit{NASA's} \textit{PolSTAR} is suited
to this application due to its sensitivity to hard X-rays.
%as is \textit{CAS's} \textit{XTP}.

\acknowledgments
AI acknowledges support from the Netherlands Organization for
Scientific Research (NWO) Veni Fellowship. AI also acknowledges useful
discussions with Chris Done. JP thanks the Academy of Finland for
support (grant 268740). We thank the anonymous referee for many useful
comments.

\bibliographystyle{apj}
\bibliography{/Users/adamingram/Dropbox/bibmaster/biblio.bib}

\vspace{-0.4cm}

\appendix

\section{A: The Kerr space-time}
\label{sec:kerr}
\noindent The non-zero entries of the Kerr metric can be expressed in Boyer-Lindquist coordinates as
\begin{equation}
g_{tt} = -\left( 1 - \frac{2r}{\Sigma} \right)~~,~~g_{t\phi} =
\frac{-2ar\sin^2\theta}{\Sigma}~~,~~g_{\phi t} = g_{t\phi}~~,~~
g_{rr}=\frac{\Sigma}{\Delta} ~~,~~g_{\theta\theta} =
\Sigma~~,~~g_{\phi\phi} = \frac{\mathcal{A}\sin^2\theta}{\Sigma},
\end{equation}
where $\Sigma \equiv r^2 + a^2\cos^2\theta$, $\Delta \equiv r^2 - 2r +
a^2$ and $\mathcal{A} \equiv (r^2+a^2)^2 - \Delta a^2
\sin^2\theta$. The 4-momentum of photons travelling in the Kerr
space-time can be expressed as (\citealt{Carter1968};
\citealt{Misner1973}; \citealt{Dovciak2004})
\begin{eqnarray}
p^t &=& \Sigma^{-1} \left[ a(l-a \sin^2\theta) + (r^2+a^2)(r^2+a^2-a
  l)/\Delta   \right] \\
p^r &=& R_{\rm sgn} \Sigma^{-1} \left[ (r^2+a^2-a l)^2 - \Delta [ (l-a)^2 + q^2 ] \right]^{1/2} \\
p^\theta &=& - \Theta_{\rm sgn} \Sigma^{-1} \left[ q^2 - \cot^2\theta (l^2 - a^2
  \sin^2\theta ) \right]^{1/2}  \\
p^\phi &=& \Sigma^{-1} \left[ l/\sin^2\theta - a + a(r^2+a^2-a l) / \Delta   \right],
\end{eqnarray}
where Carter's constants of motion, $l$ and $q^2$ are given in the
main text and the sign of the radial and polar coordinates is denoted
by $R_{\rm sgn}$ and $\Theta_{\rm sgn}$. Both of these are positive
for the case of an even number of turning points between the viewer
and the emission point and negative for an odd number.

\section{B: The flow normal}
\label{sec:tetrad}

According to the equivalence principle, we can always define a `free
falling laboratory' frame in which GR reduces to
Special Relativity. Mathematically, this can be achieved by defining a
tetrad of orthonormal unit 4-vectors. We require the flow normal
$n^\mu$ to be part of such an orthonormal tetrad in order to calculate
the emission angle from it. We calculate the $r$, $\theta$ and $\phi$
components of $n^\mu$ by transforming $\mathbf{\hat{z}_f}$ into
Boyer-Lindquist coordinates (see Appendix \ref{sec:BL}). The time-like
unit vector in the tetrad is simply $u^\mu$; i.e. the instantaneous
rest frame (e.g. \citealt{Henric2012}; \citealt{Wilkins2012}). The time-like component of
$n^\mu$ can thus be calculated by setting $n^\mu u_\mu = 0$ to give
\begin{equation}
n^t = - \frac{ \mathbf{\hat{z}_f}^d g_{d\nu} u^\nu }{ g_{t\mu} u^\mu },
\end{equation}
where $d=r,~\theta,~\phi$ but Greek letters run from $t$ to $\phi$ as
usual. Finally, we normalise to ensure $n^\mu n_\mu = 1$.

\section{C: Coordinate transforms}
\label{sec:BL}

We can convert from the BH angles $\theta$ and $\phi$ to the flow
angles $\theta_f$ and $\phi_f$ using the formulae
\begin{equation}
\cos\theta_f = \mathbf{\hat{r}} \cdot \mathbf{\hat{z}_f}
\label{eqn:thetaf}
\end{equation}
\begin{equation}
\tan\phi_f = \frac{ \mathbf{\hat{r}} \cdot \mathbf{\hat{y}_f} }{ \mathbf{\hat{r}}
  \cdot \mathbf{\hat{x}_f} }.
\label{eqn:phif}
\end{equation}
This gives
\begin{equation}
\cos\theta_f = \sin\theta\sin\beta\cos(\omega-\omega_0-\phi) +
\cos\theta\cos\beta
\label{eqn:muf}
\end{equation}
\begin{equation}
\tan\phi_f = \frac{ \sin\theta \sin(\omega-\omega_0-\phi) }
{ \cos\theta\sin\beta - \sin\theta\cos\beta \cos(\omega-\omega_0-\phi)}.
\label{eqn:tanf}
\end{equation}
The reciprocal conversion is
\begin{equation}
\cos\theta = \cos\theta_f\cos\beta + \sin\theta_f\cos\phi_f\sin\beta
\label{eqn:mu}
\end{equation}
and
\begin{equation}
\tan\phi = \frac{ -\sin\theta_f \left[ \cos\beta\sin(\omega-\omega_0)\cos\phi_f +
  \cos(\omega-\omega_0)\sin\phi_f \right] +
\cos\theta_f\sin\beta\sin(\omega-\omega_0)}
{\sin\theta_f \left[ -\cos\beta\cos(\omega-\omega_0)\cos\phi_f +
  \sin(\omega-\omega_0)\sin\phi_f \right] +
\cos\theta_f\sin\beta\cos(\omega-\omega_0)}.
\label{eqn:tan}
\end{equation}
Additionally, since $\mathbf{\hat{z}_b} =
\mathbf{\hat{z}_f}(\omega=\pi)$, we can convert to and from binary
angles using the above formulae by setting $\omega=\pi$
(i.e. replacing subscript f with subscript b and replacing $\omega$
with $\pi$).

Our calculation also requires the conversion of Cartesian coordinates
to Boyer-Lindquist coordinates. The two are related as
\begin{eqnarray}
x &=& \sqrt{r^2+a^2} \sin\theta\cos\phi \nonumber \\
y &=& \sqrt{r^2+a^2} \sin\theta\sin\phi \nonumber \\
z &=& r\cos\theta.
\label{eqn:BL}
\end{eqnarray}
To convert the coordinates of a 3-vector, $\mathbf{A}$, from a
Cartesian to a Boyer-Lindquist representation, we simply apply the
general formula 
\begin{equation}
A^{b'} = \frac{ \partial x^{b'} } { \partial x^a } A^a,
\label{eqn:transform}
\end{equation}
where $b'$ can take the values $r$, $\theta$ and $\phi$ and $a$ can
take the values $x$, $y$ and $z$. Combining Equations (\ref{eqn:BL})
gives
\begin{equation}
r^2 = \frac{1}{2} \left[ \rho^2 - a^2 + \sqrt{ (\rho^2 - a^2)^2 +
    4a^2z^2  } \right],
\label{eqn:r2}
\end{equation}
where $\rho^2 \equiv x^2 + y^2 + z^2$, along with the familiar
expressions $\cos\theta = z/r$ and $\tan\phi = y/x$. The differentials
of $r$ can be expressed as
\begin{eqnarray}
\frac{\partial r}{\partial x} &=& \sqrt{1+\left( \frac{a}{r}\right)^2}
\frac{\sin\theta\cos\phi}{2} \left[ 1 +
  \frac{r^2-a^2\cos^2\theta}{\Sigma}  \right] \nonumber \\
\frac{\partial r}{\partial y} &=& \sqrt{1+\left( \frac{a}{r}\right)^2}
\frac{\sin\theta\sin\phi}{2} \left[ 1 +
  \frac{r^2-a^2\cos^2\theta}{\Sigma}  \right] \nonumber \\
\frac{\partial r}{\partial z} &=& \frac{\cos\theta}{2}
\left[ 1 + \frac{r^2+a^2(1+\sin^2\theta)}{\Sigma}  \right].
\label{eqn:bldiffs}
\end{eqnarray}
and for $\theta$:
\begin{eqnarray}
\frac{\partial \theta}{\partial x} &=& \sqrt{1+\left( \frac{a}{r}\right)^2}
\frac{\cos\theta\cos\phi}{2r} \left[ 1 +
  \frac{r^2-a^2\cos^2\theta}{\Sigma}  \right] \nonumber \\
\frac{\partial \theta}{\partial y} &=& \sqrt{1+\left( \frac{a}{r}\right)^2}
\frac{\cos\theta\sin\phi}{2r} \left[ 1 +
  \frac{r^2-a^2\cos^2\theta}{\Sigma}  \right] \nonumber \\
\frac{\partial \theta}{\partial z} &=& - \frac{r \sin\theta}{\Sigma}.
\end{eqnarray}
Note that these expressions, as expected, reduce to the relations for
spherical polar coordinates in the Schwarzschild limit. The case of
$\sin\theta=0$ must be treated separately, although this is rather
straight forward. The differentials of $\phi$ are the same as the case
of spherical polars. These can then be substituted into Equation
(\ref{eqn:transform}) in order to convert the vector. The reciprocal
conversion, from Boyer-Lindquist to Cartesian, is far simpler
requiring the differentials of Equations (\ref{eqn:BL}) with respect
to the Boyer-Lindquist coordinates.

\end{document}

%% file: def_bibtex.tex
\def\apjs{ApJ}

\def\aap{A\&A}
\def\apjl{ApJ}